\pdfoutput=1

\documentclass[11pt]{article}

\usepackage[preprint]{acl}

\usepackage{times}
\usepackage{latexsym}
\usepackage{pifont}
\usepackage{multirow}
\usepackage{booktabs}
\usepackage{adjustbox}
\usepackage{xspace}
\usepackage{amsthm}
\usepackage[T1]{fontenc}
\usepackage[utf8]{inputenc}
\usepackage{microtype}
\usepackage{inconsolata}
\usepackage{graphicx}
\usepackage{amssymb}
\usepackage{amsmath}
\usepackage{algorithm}
\usepackage{algpseudocode}
\usepackage{subcaption}
\usepackage{wasysym}
\usepackage{colortbl}
\usepackage{xcolor}
\usepackage{textcomp}

\definecolor{lightgray}{gray}{0.9}
\definecolor{midgray}{gray}{0.6}
\definecolor{midgreen}{rgb}{0.8,1.0,0.8}

\newtheorem{theorem}{Theorem}
\newtheorem{definition}{Definition}

\newcommand{\method}{Talaria\xspace}

\title{Your Inference Request Will Become a Black Box: Confidential Inference for Cloud-based Large Language Models}

\author{
 \textbf{Chung-ju Huang\textsuperscript{1,2}},
 \textbf{Huiqiang Zhao\textsuperscript{3}},
 \textbf{Yuanpeng He\textsuperscript{1,2}},
 \textbf{Lijian Li\textsuperscript{4}},
 \\
 \textbf{Wenpin Jiao\textsuperscript{1,2}},
 \textbf{Zhi Jin\textsuperscript{1,2}},
 \textbf{Peixuan Chen\textsuperscript{3}},
 \textbf{Leye Wang\textsuperscript{1,2}}
\\
 \textsuperscript{1}Key Lab of High Confidence Software Technologies (Peking University), Ministry of Education, China,
 \\
 \textsuperscript{2}School of Computer Science, Peking University, Beijing, China,
 \\
 \textsuperscript{3}Tencent, Shenzhen, China
 \textsuperscript{4}Macau university, China
}

\newcommand{\myparagraph}[1]{\noindent \textbf{#1}\xspace}

\begin{document}
\maketitle
\begin{abstract}
The increasing reliance on cloud-hosted Large Language Models (LLMs) exposes sensitive client data, such as prompts and responses, to potential privacy breaches by service providers.
Existing approaches fail to ensure privacy, maintain model performance, and preserve computational efficiency simultaneously.
To address this challenge, we propose Talaria, a confidential inference framework that partitions the LLM pipeline to protect client data without compromising the cloud's model intellectual property or inference quality.
Talaria executes sensitive, weight-independent operations within a client-controlled Confidential Virtual Machine (CVM) while offloading weight-dependent computations to the cloud GPUs.
The interaction between these environments is secured by our Reversible Masked Outsourcing (ReMO) protocol, which uses a hybrid masking technique to reversibly obscure intermediate data before outsourcing computations.
Extensive evaluations show that Talaria can defend against state-of-the-art token inference attacks, reducing token reconstruction accuracy from over 97.5\% to an average of 1.34\%, all while being a lossless mechanism that guarantees output identical to the original model without significantly decreasing efficiency and scalability.
To the best of our knowledge, this is the first work that ensures clients’ prompts and responses remain inaccessible to the cloud, while also preserving model privacy, performance, and efficiency.
\end{abstract}

\section{Introduction}

\begin{figure}[ht]
  \centering
  \includegraphics[width=1.0\linewidth]{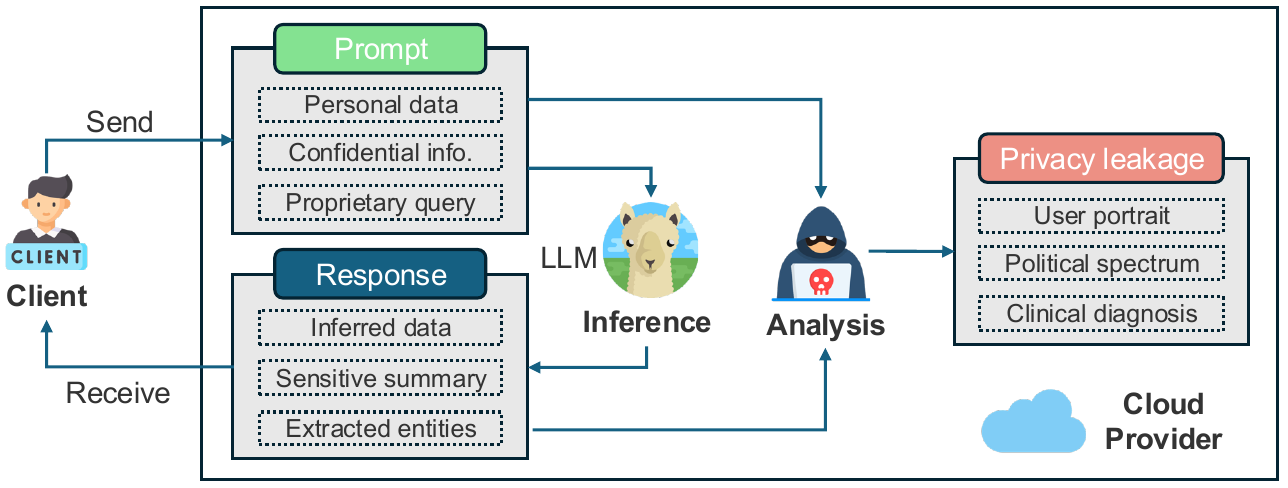}
  \caption{Privacy leakage in cloud-based LLMs.}
  \label{fig:problem}
\end{figure}

The widespread deployment of cloud-hosted Large Language Models (LLMs), such as Gemini~2.5~\cite{comanici2025gemini} and GPT~5~\cite{gpt5}, brings powerful capabilities but also an immediate privacy hazard.
Client data, including prompts and model responses, often contains personal, financial, or health information and, when processed on cloud, are exposed to the service provider (Fig.~\ref{fig:problem})~\cite{abs-2409-19134,LinHWJFZ25}.
This exposure has already chilled adoption across regulated sectors (e.g., JPMorgan, Citi, and Goldman Sachs restrictions~\cite{paul2023,brian2023}; healthcare cautions about HIPAA compliance~\cite{giles2023}), underscoring that data confidentiality becomes a deployment prerequisite for trust.
Beyond market signals, regulators likewise now expect systematic risk assessments: GDPR generally requires a Data Protection Impact Assessment (PIA) for high-risk processing, and authorities have enforced these obligations against LLM providers~\cite{gdpr}.
For example, ChatGPT was temporarily banned in Italy and later fined \texteuro15 million for training on personal data without an adequate legal basis~\cite{openai-fine}.
In parallel, U.S. FTC oversight is pushing PIA-like reviews for AI services~\cite{ftc}, and Quebec's Law~25 makes PIAs mandatory with significant penalties~\cite{Quebec}.
Consequently, any trustworthy cloud-based LLM solution must provide end-to-end safeguards for prompts (and responses) to demonstrate PIA compliance.
%

\myparagraph{Challenges.}
While ensuring prompt-response confidentiality, the primary challenge is satisfying three conflicting requirements, often termed the ``impossible trinity''~\cite{abs-2409-19134,abs-2412-10652}:
\ding{172} \textit{Model Privacy}, which requires safeguarding client data without compromising the confidentiality of cloud-hosted models;
\ding{173} \textit{Model Performance}, which requires output invariance, i.e., that new mechanisms do not alter the model's outputs;
\ding{174} \textit{Model Efficiency}, which requires that privacy mechanisms introduce only modest computational overhead.
\textit{This paper aims to realize a confidential inference mechanism that prevents the cloud from obtaining the client's prompt and response while balancing the impossible trinity of model privacy, performance, and efficiency.}


\myparagraph{Status Quo.}
Current approaches to private LLM inference uniformly fail to balance this trinity (Tbl.~\ref{tbl:status_quo}).
Data-centric methods, including differential privacy~\cite{WuPWM24,abs-2505-05922} and anonymization~\cite{abs-2309-03057,abs-2406-01394}, compromise performance by altering inputs, which violates output invariance.
Homomorphic encryption-based mechanisms~\cite{abs-2305-18396,HuangLHD22} provide provable security, but the large computational overhead and the need to access model parameters limit their widespread application in LLMs.
Confidential Virtual Machines (CVMs) with GPU support have been proposed as a path forward, promising greater efficiency by leveraging secure hardware enclaves.
While CVMs can effectively isolate the runtime environment, existing implementations remain vulnerable; they do not prevent leakage of model parameters and the response, thus failing to provide comprehensive privacy guarantees~\cite{abs-2409-19134}.

\begin{figure}[ht]
  \centering
  \includegraphics[width=1.0\linewidth]{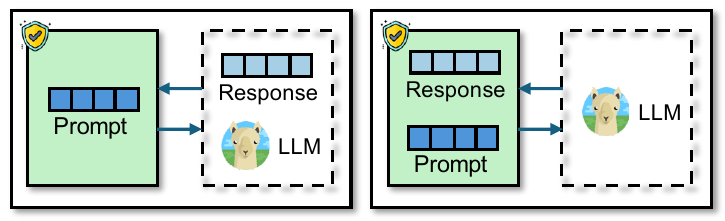}
  \caption{The left figure shows the existing CVM-based methods, which isolates the prompt in the CVM but exposes the response. The right figure shows our method, which isolates both prompt and response in CVM.}
  \label{fig:comparison}
\end{figure}

\myparagraph{Our Solution.}
To address these limitations, we propose \method~\footnote{In Greek mythology, Hermes wore the winged sandals \method, which allowed him to travel freely between the divine realm and the mortal world. Our method enables client data to traverse securely and efficiently between secure and non-secure zones.}, a novel CVM-based confidential inference paradigm that protects client prompts and responses by balancing model privacy, performance, and efficiency.
The difference between \method and the existing CVM-based methods is shown in Fig.~\ref{fig:comparison}.
\method uniquely partitions the LLM inference pipeline:
sensitive operations involving client data are executed within a secure, client-controlled cloud CVM, while computationally intensive, weight-dependent operations are offloaded to the public cloud GPU.
This Confidential Partitioned Inference (CPI) interaction is secured by our novel Reversible Masked Outsourcing (ReMO) protocol, which provides two-way protection by preventing the cloud from seeing client data and the client from inferring model weights.
Empirically, \method thwarts state-of-the-art attacks, reducing token reconstruction accuracy from over 97.5\% to just 1.34\%.
It achieves this robust security losslessly--guaranteeing bit-for-bit identical output to the original model--while demonstrating superior efficiency over existing CVM-based secure inference methods.

\myparagraph{Our Contributions.}
The contributions of our work can be summarized as follows:
\begin{itemize}
    \item To the best of our knowledge, this is the first work that protects client's prompts and responses in cloud-based LLM inference.
    \item \method provides strong data privacy protection while maintaining a balance between model privacy, performance, and efficiency.
\end{itemize}

\section{Related Work}

\begin{table}
  \centering
  \adjustbox{max width=1.0\linewidth}{
  \begin{tabular}{cccccc}
    \toprule
    \multirow{2}{*}{\textbf{Solution}} & \multicolumn{3}{c}{\textbf{Privacy}} & \textbf{Performance} & \multirow{2}{*}{\textbf{Efficiency}} \\ 
    \cmidrule(lr){2-4} \cmidrule(lr){5-5}
    & Prompt & Response & Model & Output variance & \\
    \midrule
    DP   & \LEFTcircle & \Circle & \CIRCLE     & \Circle & \LEFTcircle \\
    HE   & \CIRCLE     & \Circle & \Circle     & \Circle & \Circle \\
    CVM  & \CIRCLE     & \Circle & \Circle     & \CIRCLE & \LEFTcircle \\
    Ours & \CIRCLE     & \CIRCLE & \LEFTcircle & \CIRCLE & \LEFTcircle \\
    \bottomrule
  \end{tabular}
  }
  \caption{\label{tbl:status_quo}
  Comparison of existing solutions (\CIRCLE: Fully support, \Circle: Not support, \LEFTcircle: Partial support).}
\end{table}

\myparagraph{DP-based methods.}
Differential privacy~\cite{Dwork06} offers an efficient way to protect client prompts by perturbing the inputs, but this often comes at the cost of performance.
DP-based techniques inject noise into token distributions, embeddings, or use paraphrasing to obscure original inputs~\cite{abs-2505-05922,WuPWM24,HongWZL0W24,LinYMZH0WLCD025}.
While highly efficient and not requiring changes to the backend model, this noise injection inevitably creates a trade-off between privacy and performance, as it can degrade the quality of downstream tasks.


\myparagraph{HE-based methods.}
Homomorphic Encryption, often used with secure multi-party computation, allows for direct computation on encrypted data, promising strong, provable security for LLM inference without degrading output quality~\cite{abs-2305-18396,HuangLHD22}.
However, the practical application of these methods to large-scale cloud-based LLMs is hindered by two major obstacles: prohibitive computational overhead that violates efficiency, and the need for model access, which compromises the model confidentiality.

\myparagraph{CVM-based methods.}
CVMs provide a hardware-based TEE to isolate client data, but practical implementation for LLMs remains challenging.
The naive approach of running one full LLM instance per client CVM, while secure~\cite{Hanzlik0G0A0F21,LeeLPLLLXXZS19}, is unscalable and economically impractical due to the immense memory requirements of LLMs.
Hybrid solutions attempt to solve this by processing prompts inside the CVM while offloading token generation to the cloud~\cite{abs-2409-19134}.
This improves efficiency but critically undermines the security model: it violates model confidentiality by exposing the model to the client and compromises client privacy by revealing the plaintext response to the cloud provider.

\section{Threat Model}

We consider a typical two-party cloud inference setting with:
a cloud provider holds LLMs, and a client holds privacy-sensitive prompts and receives model responses.

\myparagraph{Cloud.}
Consistent with related work~\cite{abs-2409-19134,abs-2505-05922}, we assume the cloud provider operates under the \textit{honest-but-curious} threat model.
The cloud is \textit{honest} as it is economically motivated to follow the inference protocol faithfully; its business model relies on customer trust, which would be destroyed by any detected malicious tampering~\cite{SharmaC21,XueCLHH18}.
Simultaneously, the cloud is \textit{curious}, meaning it may leverage its extensive computational resources to analyze prompts and responses, aiming to reconstruct sensitive client information or user attributes via inversion attacks~\cite{abs-2503-09291,0004ZWXYLZ25}.


\myparagraph{Client.}
The client is also \emph{honest-but-curious}: it adheres to the protocol to obtain inference results while attempting to learn the cloud's proprietary parameters (e.g., reverse-engineering model weights) from any messages it legitimately receives.

\myparagraph{Trusted Computing Base (CVM).}
Our security model is based on the availability of GPU-accelerated CVMs.
This assumption is practical, as major cloud platforms like Amazon Web Services~\cite{amazon}, Microsoft Azure~\cite{azure}, and Google Cloud~\cite{googlecloud} already provide such offerings.
The client can verify the integrity of a CVM through remote attestation.
Once verified, the CVM is considered a trusted extension of the client, allowing them to form a single logical party and securely share secrets.
The CVM's runtime state is cryptographically isolated from the cloud; the provider cannot directly access plaintext data inside the CVM.
While the cloud cannot read the CVM's runtime data, it retains control over how the CVM is provisioned.
Following industry practice~\cite{AzureAI,Nvidia-llm-security}, the provider may enforce a minimal, pre-approved software stack and capability set to reduce misuse and to narrow the attack surface (e.g., blocking privileged or abuse-prone instructions and interfaces~\cite{SchwarzR24}, or enforcing launch control policies that only allow signed enclave binaries~\cite{SchwarzWGMM20}).
These controls prevent a curious client from turning the CVM into an oracle for model extraction or systems introspection, without granting the cloud visibility into the CVM's plaintext.
%


\myparagraph{Out of scope.}
Following standard CVM practice~\cite{abs-2409-19134,Hanzlik0G0A0F21}, we exclude microarchitectural and side-channel attacks, physical attacks, supply-chain compromises, and denial-of-service.
We also do not consider a fully malicious cloud that arbitrarily deviates from the protocol (see Sec.~\ref{sec:limitation} for discussion).

\section{Methodology}

\subsection{Design Insight}

\begin{figure}[ht]
  \centering
  \includegraphics[width=1.0\linewidth]{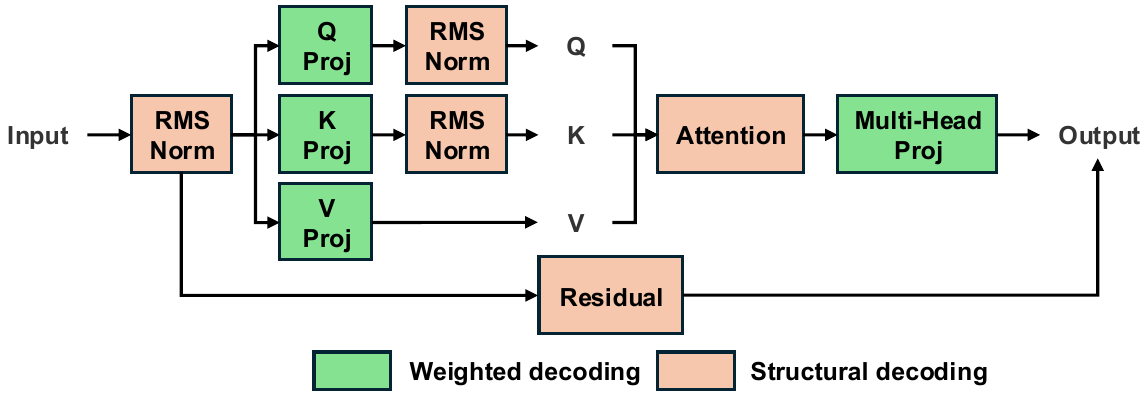}
  \caption {
  Using a Qwen~\cite{qwen3technicalreport} attention layer as an example, we split inference into two parts.
  The green box (weighted decoding) performs linear projections that apply model weights $W$ to inputs.
  The yellow box (structural decoding) comprises weight-free ops--RMSNorm~\cite{ZhangS19a}, attention score/softmax~\cite{VaswaniSPUJGKP17}, and residual connections.
  We run green-box ops on public GPUs and yellow-box ops on cloud CVMs to minimize exposure of confidential weights.}
  \label{fig:illustration}
\end{figure}

\method enables Confidential Partitioned Inference (CPI) by decoupling computations that require proprietary parameters from computations that touch client plaintext. Concretely, we partition the standard LLM layer into:
\begin{itemize}
    \item \textbf{Weighted (weight-dependent) decoding:} linear projections and other ops that multiply by model weights $W$.
    \item \textbf{Structural (weight-free) decoding:} normalization, attention-score/softmax, residuals, and sampling that do not require direct access to $W$.
\end{itemize}
We run the first category on the cloud GPU (holding $W$) and the second inside the CVM, preventing the cloud from learning client data and the client from learning $W$ (Fig.~\ref{fig:illustration}).

\paragraph{Notations.}
We use $E\in\mathbb{R}^{n\times d}$ for a layer's intermediate embeddings (sequence length $n$, hidden dimension $d$), $W\in\mathbb{R}^{d\times d}$ for the relevant weight matrix, $\hat{E}$ and $\hat{O}$ for masked inputs/outputs, $O=EW$ for the true weighted result, and $X$ for tokens.
We list all the notations in Appx.~\ref{appx:notation}.

\subsection{System Overview}
\label{sec:overview}

\begin{figure}[ht]
  \includegraphics[width=1.0\linewidth]{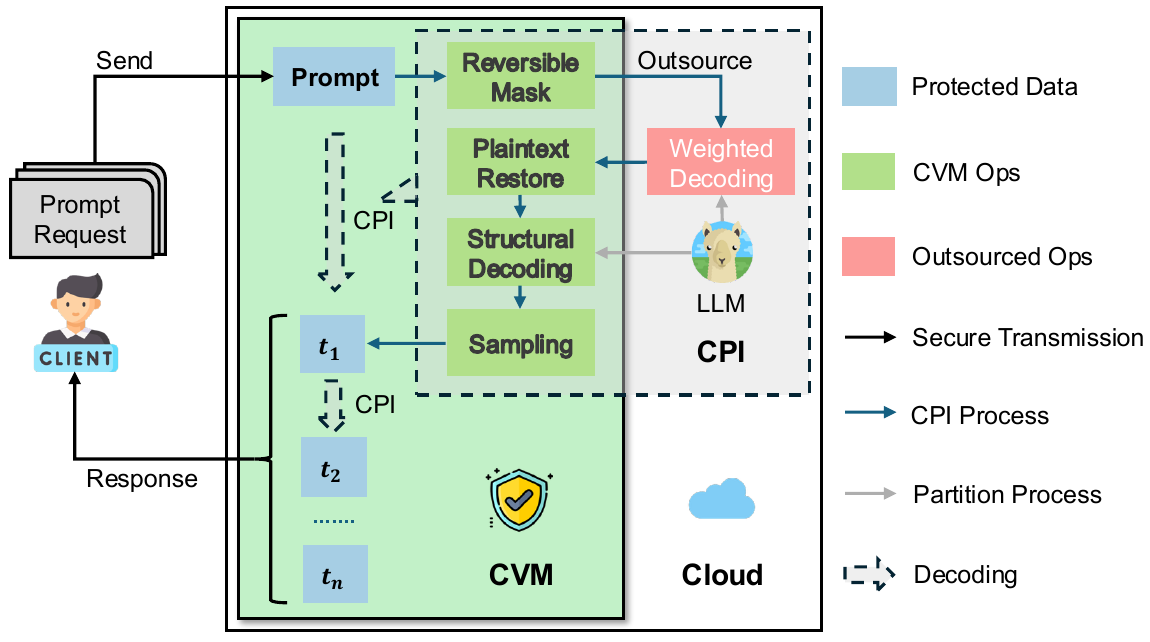}
  \caption{
  The overview of \method.
  The client first transmits the prompt to the CVM through a secure channel.
  For each new token generated, we use CPI for secure collaborative inference.
  When all tokens are generated, the CVM sends the response back to the client confidentially.
  All private data is isolated in the CVM and is invisible to the cloud.
  }
  \label{fig:overview}
\end{figure}

The confidential inference process of \method is shown in Fig.~\ref{fig:overview}.

\subsubsection{One-time setup.}
The client remotely attests a CVM and establishes a secure channel. The provider loads the model on a public GPU.
We then precompute a reusable restoration pool for later recovery (Sec.~\ref{sec:hybrid-masking}).

\myparagraph{Model partitioning.}
Each layer is split so that linear ops (e.g., Q/K/V and output projections in attention; MLP projections) execute on the unprotected cloud GPU, while structural decoding ops execute inside the CVM.

\myparagraph{Restoration pool preparation.}
The CVM generates a public base $M_{\mathrm{pub}}\in\mathbb{R}^{m\times d}$ with $m<d$, sends it to the GPU, and receives $R_{\mathrm{pub}}=M_{\mathrm{pub}} W$ back. $R_{\mathrm{pub}}$ will let the CVM later undo outsourced multiplications without ever obtaining $W$.

\subsubsection{Iterative inference.}
For each decoding step, the client’s prompt $X$ is decrypted inside the CVM and embedded to $E$.
Before outsourcing any weight multiplication, the CVM masks $E$ (Sec.~\ref{sec:remo}), sends only the masked tensor to the unprotected GPU for weighted decoding, and then locally restores the exact plaintext result and proceeds with structural decoding.
The private KV cache remains in the CVM.
The process repeats until $\mathsf{EOS}$; see pseudocode in Appx.~\ref{appx:code}.

\subsection{Reversible Masked Outsourcing (ReMO)}
\label{sec:remo}

Naively outsourcing $EW$ reveals $E$ to the cloud.
ReMO prevents this with a reversible additive mask.
Given $E\in\mathbb{R}^{n\times d}$ and $W\in\mathbb{R}^{d\times d}$, the CVM samples a secret mask $M\in\mathbb{R}^{n\times d}$ and forms
\[
\hat{E} = E + M.
\]
Only $\hat{E}$ is sent to the GPU, which computes
\[
\hat{O} = \hat{E} W = EW + MW.
\]
To recover the true result $O=EW$ inside the CVM, we must subtract the restoration term $R=MW$.
The core challenge is that $R$ depends jointly on the CVM secret $M$ and the cloud secret $W$.
Our protocol ensures neither secret ever leaves its trust boundary while still enabling exact restoration.

\paragraph{Correctness (output invariance).}
ReMO is algebraically lossless: after restoration (Sec.~\ref{sec:hybrid-masking}), the CVM obtains exactly $O=EW$.
Hence, under a fixed sampler, responses with and without \method are bitwise identical (validated in Sec.~\ref{sec:eval-invariance}).

\subsection{Hybrid Masking Protocol}
\label{sec:hybrid-masking}

Hybrid masking separates mask generation from mask application so that the CVM can compute $R=MW$ without $W$ and the cloud can process masked data without $M$.

\paragraph{Public base (setup).}
The CVM samples a public base $M_{\mathrm{pub}}\in\mathbb{R}^{m\times d}$, $m<d$, and sends it to the GPU. The GPU returns the pre-multiplied pool
\[
R_{\mathrm{pub}} \triangleq M_{\mathrm{pub}} W,
\]
which the CVM stores locally.
$M_{\mathrm{pub}}$ is non-sensitive; $R_{\mathrm{pub}}$ is safe to keep in the CVM because, with $m<d$, it reveals an underdetermined sketch of $W$ (formalized in Sec.~\ref{sec:model-confidentiality}).

\paragraph{Ephemeral private mixing (per token).}
For each decoding step with sequence length $n$, the CVM samples a fresh, private $M_{\mathrm{pvt}}\in\mathbb{R}^{n\times m}$ (never leaves the CVM) and derives
\[
M = M_{\mathrm{pvt}} M_{\mathrm{pub}}.
\]
It forms $\hat{E}=E+M$ and outsources $\hat{E}$. After receiving $\hat{O}=\hat{E}W$, the CVM restores the plaintext result using only local secrets:
\begin{equation}
\label{eq:recover}
O \;=\; \hat{O} - M W \;=\; \hat{O} - M_{\mathrm{pvt}} R_{\mathrm{pub}}.
\end{equation}

\paragraph{Security hooks for later analysis.}
Eq.~\eqref{eq:recover} keeps $M$ and $W$ in separate trust boundaries, supporting:
\begin{itemize}
    \item \emph{Data confidentiality (Sec.~\ref{sec:data-confidentality}):} Each token uses a fresh, full-space mask (instantiated via a pseudorandom generator (PRG) inside the CVM), matching Def.~\ref{def:ci}.Additive masking gives the quantitative bound in Thm.~\ref{thm:its}.
    \item \emph{Model confidentiality (Sec.~\ref{sec:model-confidentiality}):} With $m<d$, $R_{\mathrm{pub}}=M_{\mathrm{pub}}W$ leaves a nontrivial right-kernel, so recovering $W$ from $(M_{\mathrm{pub}},R_{\mathrm{pub}})$ is information-theoretically impossible.
\end{itemize}


\paragraph{Practical notes.}
$R_{\mathrm{pub}}$ is prepared once per operation $W$ and reused across tokens; $M_{\mathrm{pvt}}$ is sampled per token to ensure freshness.
The KV cache is private and persists only inside the CVM.

\subsection{Security Challenge and Threat Focus}
\label{sec:challenge}

Outsourcing weighted decoding naively exposes $E$ to a cloud adversary $\mathcal{A}$, enabling prompt/response inversion $X'\!\leftarrow\!\mathcal{A}(E)$ with $X'\!\approx\!X$.
ReMO with hybrid masking removes this leakage by ensuring the GPU only ever sees masked tensors and never receives enough information to infer $M$ (or undo it), while the CVM never obtains $W$.

\section{Privacy Guarantee Analysis}

\subsection{Data Confidentiality}
\label{sec:data-confidentality}
\noindent\textbf{Adversary and domain.}
We protect client prompts and responses against an ``honest-but-curious'' cloud that may adaptively query the protocol and analyze all public transcripts.
We treat intermediate embeddings $E\in\mathbb{R}^{n\times d}$ under a fixed, deterministic quantization into a finite ring (e.g., $\mathbb{Z}_{2^k}$) so that masking and indistinguishability are defined over a finite domain.
All results extend to reals via standard discretization arguments.

\myparagraph{Mask generation and freshness.}
For each decoding step (per token), the CVM samples a fresh, ephemeral private matrix $M_{\mathrm{pvt}}\!\in\!\mathbb{R}^{n\times m}$ and forms an additive mask $M=M_{\mathrm{pvt}} M_{\mathrm{pub}}$ using a public base $M_{\mathrm{pub}}\!\in\!\mathbb{R}^{m\times d}$ established during setup.
The mask is applied entrywise: $\hat{E}=E+M$.
Freshness (single-use) and full-space randomness ensure the masked distribution does not concentrate on any known low-dimensional subspace.

\begin{definition}[Computational indistinguishability]
\label{def:ci}
Let $\mathcal{A}$ be any PPT adversary with access to all public auxiliary information $Z$ (including protocol transcripts and $M_{\mathrm{pub}}$).
For any two inputs $E_1,E_2\in\mathbb{R}^{n\times d}$ with corresponding masked versions $\hat{E}_1,\hat{E}_2$, the distinguishing advantage is
\begin{align}
\bigl|\,\Pr[\mathcal{A}(\hat{E}_1,Z)=1]&-\Pr[\mathcal{A}(\hat{E}_2,Z)=1]\,\bigr|\notag \\
&\le \varepsilon(\kappa,q),
\end{align}
where $\kappa$ is the security parameter of PRG used to derive masks and $q$ is the number of masked samples observed.
If $\varepsilon$ is negligible in $\kappa$ for any polynomially bounded $q$, the scheme is privacy-preserving.
\end{definition}

\noindent\textit{Instantiation in \method.}
Within the CVM, masks are derived from a PRG keyed by a secret seed of length $\kappa$; conditioned on $Z$, the mask distribution is computationally indistinguishable from uniform over the ring (Appx.~\ref{appx:proof-def}).
Because a fresh mask is used per token, the cloud cannot correlate observations across steps to reduce uncertainty.

\myparagraph{Quantitative information-theoretic bound.}
Beyond computational security, the masking admits an information-theoretic guarantee that depends only on signal magnitude versus mask range.

\begin{theorem}[Information-theoretic bound]
\label{thm:its}
Let $e=\mathrm{vec}(E)\in\mathbb{R}^{N}$ with $N=nd$.
Suppose the mask has i.i.d.\ coordinates $m_i\sim\mathrm{Unif}[-\lambda/2,\lambda/2]$ and $\hat{e}_b=e_b+m$ for $b\in\{1,2\}$.
Then the optimal adversary (with unbounded computation) that observes $\hat{e}$ satisfies
\begin{align}
\Pr[\text{correctly distinguish }E_1\text{ from }E_2\mid \hat{e}]\ \notag \\\le\ \tfrac{1}{2}\ +\ \tfrac{1}{2}\,\min\!\left\{ \frac{\|e_1-e_2\|_1}{\lambda},\, 1 \right\}.
\end{align}
\end{theorem}

\noindent\textit{Implications for parameterization.}
The distinguishing advantage scales as (effective) signal over mask range. Choosing $\lambda$ so that $\|e_1-e_2\|_1/\lambda$ is negligible for inputs drawn from the task distribution makes the optimal success probability arbitrarily close to $1/2$.
Practically, one can upper-bound $\|e_1-e_2\|_1$ via sensitivity estimates of the embedding map and set $\lambda$ with a safety margin to cover worst-case layers/tokens (Appx.~\ref{appx:proof-thm}).

\subsection{Model Confidentiality}
\label{sec:model-confidentiality}
\noindent\textbf{Goal.}
Prevent a curious client from recovering proprietary weights $W\in\mathbb{R}^{d\times d}$.

\myparagraph{Mathematical isolation.}
The client observes only the linearly transformed $R_{\mathrm{pub}}=M_{\mathrm{pub}}W$ with $M_{\mathrm{pub}}\in\mathbb{R}^{m\times d}$ and $m<d$.
Since $\operatorname{rank}(M_{\mathrm{pub}})\le m<d$, the right null space is nontrivial, $\dim(\ker(M_{\mathrm{pub}}))\ge d-m>0$, and
\begin{align}
    &\{W'\!:\,M_{\mathrm{pub}}W'=R_{\mathrm{pub}}\} =\\ 
    W_0 + \{&Z:\ \text{each column of }Z\in\ker(M_{\mathrm{pub}})\}. \notag
\end{align}
Thus the feasible set is an affine subspace of positive dimension (infinitely many candidates), making unique recovery of $W$ information-theoretically impossible. Here $\ker(M)=\{v\in\mathbb{R}^d:\,Mv=0\}$.

\myparagraph{System-level hardening.}
We enforce remote attestation of the CVM image and restrict its capabilities to the minimal \method logic: no debug hooks, no general linear-algebra toolkits exposing arbitrary solves, and no interfaces that could be repurposed to probe $R_{\mathrm{pub}}$ beyond its intended use.
Attestation binds code and configuration to a hash verified by the cloud, preventing a curious client from modifying the CVM to mount offline model-recovery oracles.
See Appx.~\ref{appx:proof-model} for formal proofs.

\section{Evaluations}

\subsection{Evaluation Setup}

All evaluations are conducted on a cloud node with 8 Nvidia H800 with 80GB of memory, 10 AMD EPYC processor cores, and 6400 GB of system memory.
We enable security features such as AMD SEV-SNP~\cite{amdsev} and NVIDIA GPU CC~\cite{nvidiacc} for evaluations that require confidential computing.
Detailed evaluation setup (datasets, metrics, etc.) can be found in Appx.~\ref{appx:setup}.

\myparagraph{Datasets and models.}
We evaluate \method on four diverse datasets: Midjourney prompts~\cite{midjourney-prompts}, WikiText-2~\cite{MerityX0S17}, Patient-notes~\cite{patientnotes}, and GPT-OSS20B-samples~\cite{gpt-oss20b-samples}.
To simulate our threat model, we use 20\% of each dataset as private client prompts and allow the remaining 80\% to be used for training attack models.
Our experiments leverage models from the Llama3 (1B, 8B, 70B)~\cite{llama3} and Qwen3 (0.6B, 8B, 32B)~\cite{qwen3} families.
By default, we use Qwen3-8B for all evaluations unless stated otherwise.

\myparagraph{Metrics.}
For protection evaluation, we use Token Reconstruction Accuracy (TRA) and Cosine Semantic Similarity (CSS).
The higher the TRA and CSS, the higher the similarity between the two texts, which can reflect the higher accuracy of the token reconstruction attack.
For efficiency evaluation, we use end-to-end latency and Time to First Token (TTFT).

\myparagraph{Attacks.}
We validate our defense against three SOTA reconstruction attacks targeting intermediate data:
TokenInfer~\cite{abs-2503-09291} (classification-based), TokenInv~\cite{0004ZWXYLZ25} (optimization-based), and SentenceInv~\cite{GuKRVM23} (generation-based).
By default, we use TokenInfer for most evaluations.
According to TokenInfer, the reconstruction performance decreases as the number of layers increases.
Therefore, in order to maximize the attack effect, we use the outputs of the first attention layer as the attack object.


\myparagraph{Baselines.}
We first use \textit{No Protection} and \textit{Full Isolation} as the baseline, where the former is normal unprotected cloud inference and the latter is where each client loads a copy of the complete LLM into CVM for inference.
Then we also use two SOTA defense mechanisms as baselines: SPD~\cite{abs-2409-19134}, Cape~\cite{abs-2505-05922}.

\subsection{Protection Evaluations}
We assess data privacy protection along two dimensions: input prompts and output responses.
To facilitate comparison of metrics, we set the maximum number of output tokens per request to 100.
Additional evaluations can be found at Appx.~\ref{appx:additional_protection_eval}.

\begin{table*}[t]
  \centering
  \adjustbox{max width=1.0\linewidth}{
  \begin{tabular}{clcccccccc}
    \hline
    \textbf{Attack Target} & \textbf{Methods}        & \multicolumn{2}{c}{\textbf{Midjourney}} & \multicolumn{2}{c}{\textbf{WikiText2}} & \multicolumn{2}{c}{\textbf{Patient}} & \multicolumn{2}{c}{\textbf{GPT-samples}} \\
    \hline
    \multirow{3}{*}{Prompt} 
    & \textit{No Protection} & \textbf{96.87\%} & \textcolor{midgray}{0.9352} & \textbf{97.21\%} & \textcolor{midgray}{0.9444} & \textbf{98.64\%} & \textcolor{midgray}{0.9834} & \textbf{97.53\%} & \textcolor{midgray}{0.9421} \\
    & Cape                   & \textbf{85.25\%} & \textcolor{midgray}{0.8824} & \textbf{84.55\%} & \textcolor{midgray}{0.8984} & \textbf{84.12\%} & \textcolor{midgray}{0.9023} & \textbf{85.89\%} & \textcolor{midgray}{0.8334} \\
    & \method                & \cellcolor{lightgray}\textbf{1.24\%} & \cellcolor{lightgray}\textcolor{midgray}{0.4876} & 
                               \cellcolor{lightgray}\textbf{0.47\%} & \cellcolor{lightgray}\textcolor{midgray}{0.4299} & 
                               \cellcolor{lightgray}\textbf{1.53\%} & \cellcolor{lightgray}\textcolor{midgray}{0.5012} & 
                               \cellcolor{lightgray}\textbf{2.11\%} & \cellcolor{lightgray}\textcolor{midgray}{0.4535} \\
    \hline
    \multirow{4}{*}{Response} 
    & \textit{No Protection} & \textbf{97.53\%} & \textcolor{midgray}{0.9326} & \textbf{97.54\%} & \textcolor{midgray}{0.9401} & \textbf{97.28\%} & \textcolor{midgray}{0.9885} & \textbf{97.68\%} & \textcolor{midgray}{0.9456} \\
    & Cape                   & \textbf{85.53\%} & \textcolor{midgray}{0.8921} & \textbf{84.65\%} & \textcolor{midgray}{0.8906} & \textbf{84.64\%} & \textcolor{midgray}{0.9005} & \textbf{86.10\%} & \textcolor{midgray}{0.8389} \\
    & SPD                    & \textbf{97.16\%} & \textcolor{midgray}{0.9344} & \textbf{96.87\%} & \textcolor{midgray}{0.9375} & \textbf{98.76\%} & \textcolor{midgray}{0.9903} & \textbf{97.24\%} & \textcolor{midgray}{0.9413} \\
    & \method                & \cellcolor{lightgray}\textbf{1.08\%} & \cellcolor{lightgray}\textcolor{midgray}{0.4724} & 
                               \cellcolor{lightgray}\textbf{0.38\%} & \cellcolor{lightgray}\textcolor{midgray}{0.4242} & 
                               \cellcolor{lightgray}\textbf{1.58\%} & \cellcolor{lightgray}\textcolor{midgray}{0.4980} & 
                               \cellcolor{lightgray}\textbf{1.97\%} & \cellcolor{lightgray}\textcolor{midgray}{0.4463} \\
    \hline
  \end{tabular}
  }
  \caption{\label{tab:protection_main_eval}Effectiveness of different methods in defending against TokenInfer attack. 
  Bold: TRA. Gray: CSS. Shaded cells highlight the best protection.}
\end{table*}

\myparagraph{Comparison with baselines.}
We evaluate the protection performance of \method against several baselines, with results summarized in Tbl.~\ref{tab:protection_main_eval}.
\textit{No Protection} highlights the severe privacy risk, as an adversary can infer both prompts and responses with near-perfect accuracy (avg. TRA > 97.5\% and avg. CSS > 0.95).
This demonstrates that without protecting the intermediate data, the cloud can easily infer the private prompt and response.
While existing methods like Cape offer limited mitigation, they remain largely vulnerable.
Another baseline, SPD, effectively protects prompts by processing the prefill phase within the CVM.
However, this approach has a critical flaw: it exposes confidential model weights to the client.
Furthermore, SPD leaves the generated response completely unprotected, as intermediate data during the decoding phase is transmitted in plaintext, leading to similarly high attack success rates (avg. TRA: 97.51\%, avg. CSS: 0.9509).
In stark contrast, \method provides robust protection across the board.
It slashes the average TRA for prompt inference from 97.56\% to just 1.34\% and for response inference from 97.51\% to 1.25\%.
Similarly, CSS plummets from over 0.95 to under 0.47.
These results demonstrate that \method effectively defends inference attacks, rendering the recovery of private tokens nearly impossible.

\myparagraph{Different LLMs.}
To assess the generalizability of our approach, we evaluate \method across a diverse set of models with varying architectures and parameter scales.
As shown in Tbl.~\ref{tab:protection_model_eval}, \method consistently delivers strong protection for both client prompts and model responses, demonstrating its robustness and broad applicability.

\begin{table}[ht]
  \centering
  \adjustbox{max width=1.0\linewidth}{
  \begin{tabular}{clccc}
    \toprule
    \textbf{Attack Target} & \multicolumn{2}{c}{\textbf{Models}} & \textbf{TRA} & \textbf{CSS} \\
    \midrule
    \multirow{6}{*}{Prompt}   & \multirow{3}{*}{Qwen3}  & 0.6B & 1.56\% & 0.4573 \\
                                                      & & 8B   & 1.24\% & 0.4876 \\
                                                      & & 32B  & 1.43\% & 0.4753 \\ \cmidrule{2-5}
                              & \multirow{3}{*}{Llama3} & 1B   & 1.46\% & 0.4323 \\
                                                      & & 8B   & 1.39\% & 0.4632 \\
                                                      & & 70B  & 1.57\% & 0.4553 \\
    \midrule
    \multirow{6}{*}{Response} & \multirow{3}{*}{Qwen3}  & 0.6B & 1.34\% & 0.4326 \\
                                                      & & 8B   & 1.08\% & 0.4724 \\
                                                      & & 32B  & 1.21\% & 0.4683 \\ \cmidrule{2-5}
                              & \multirow{3}{*}{Llama3} & 1B   & 0.89\% & 0.4521 \\
                                                      & & 8B   & 1.15\% & 0.4646 \\
                                                      & & 70B  & 0.78\% & 0.4328 \\
    \bottomrule
  \end{tabular}
  }
  \caption{\label{tab:protection_model_eval}The effectiveness of \method in defending against TokenInfer attack under different model structures and parameters (Midjourney dataset).}
\end{table}

\myparagraph{Different Attacks.}
To rigorously assess its security, we evaluate \method against three SOTA attacks: TokenInfer, TokenInv, and SentenceInv.
Against TokenInfer, the most potent of the three, \method proves exceptionally effective, reducing the average TRA to a mere 1.16\% and the average CSS to 0.48.
Our approach is similarly successful in thwarting TokenInv.
Furthermore, \method effectively defends against SentenceInv, a prompt-only reconstruction attack for which TRA is not an applicable metric.
These results demonstrate that \method provides comprehensive protection against a diverse range of sophisticated attacks.

\begin{table}[ht]
  \centering
  \adjustbox{max width=1.0\linewidth}{
  \begin{tabular}{clcc}
    \toprule
    \textbf{Attack Target} & \textbf{Attacks} & \textbf{TRA} & \textbf{CSS} \\
    \hline
    \multirow{3}{*}{Prompt}   & TokenInfer  & 1.24\% & 0.4876 \\
                              & TokenInv    & 0.45\% & 0.3893 \\
                              & SentenceInv & -      & 0.4435 \\ \hline
    \multirow{3}{*}{Response} & TokenInfer  & 1.08\% & 0.4724 \\
                              & TokenInv    & 0.56\% & 0.3954 \\
                              & SentenceInv & -      & -      \\ 
    \bottomrule
  \end{tabular}
  }
  \caption{\label{tab:protection_attack_eval}The effectiveness of \method in defending against different attacks (Midjourney dataset).}
\end{table}


\subsection{Efficiency Evaluations}
Our evaluation setup includes 1 to 32 clients, each of whom randomly selects 10 prompts from the Midjourney dataset.
For ease of comparison, we fix the number of output tokens to 100.
Finally, we calculate the average efficiency metrics.

\begin{figure*}[!htbp]
    \centering
    \begin{subfigure}{0.32\linewidth}
        \centering
        \includegraphics[width=\linewidth]{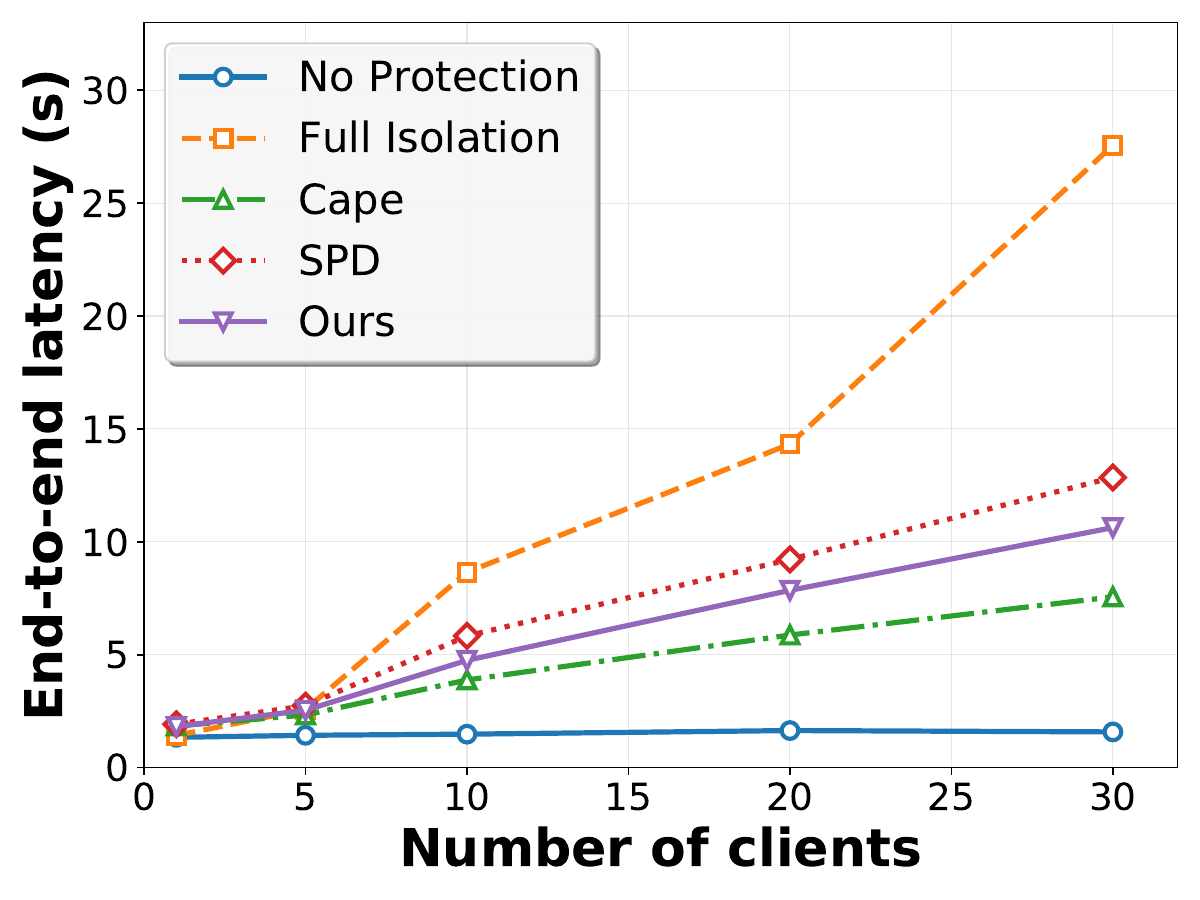}
        \caption{End-to-end latency with varying number of clients.}
        \label{fig:num_of_clients}
    \end{subfigure}
    \hfill
    \begin{subfigure}{0.32\linewidth}
        \centering
        \includegraphics[width=\linewidth]{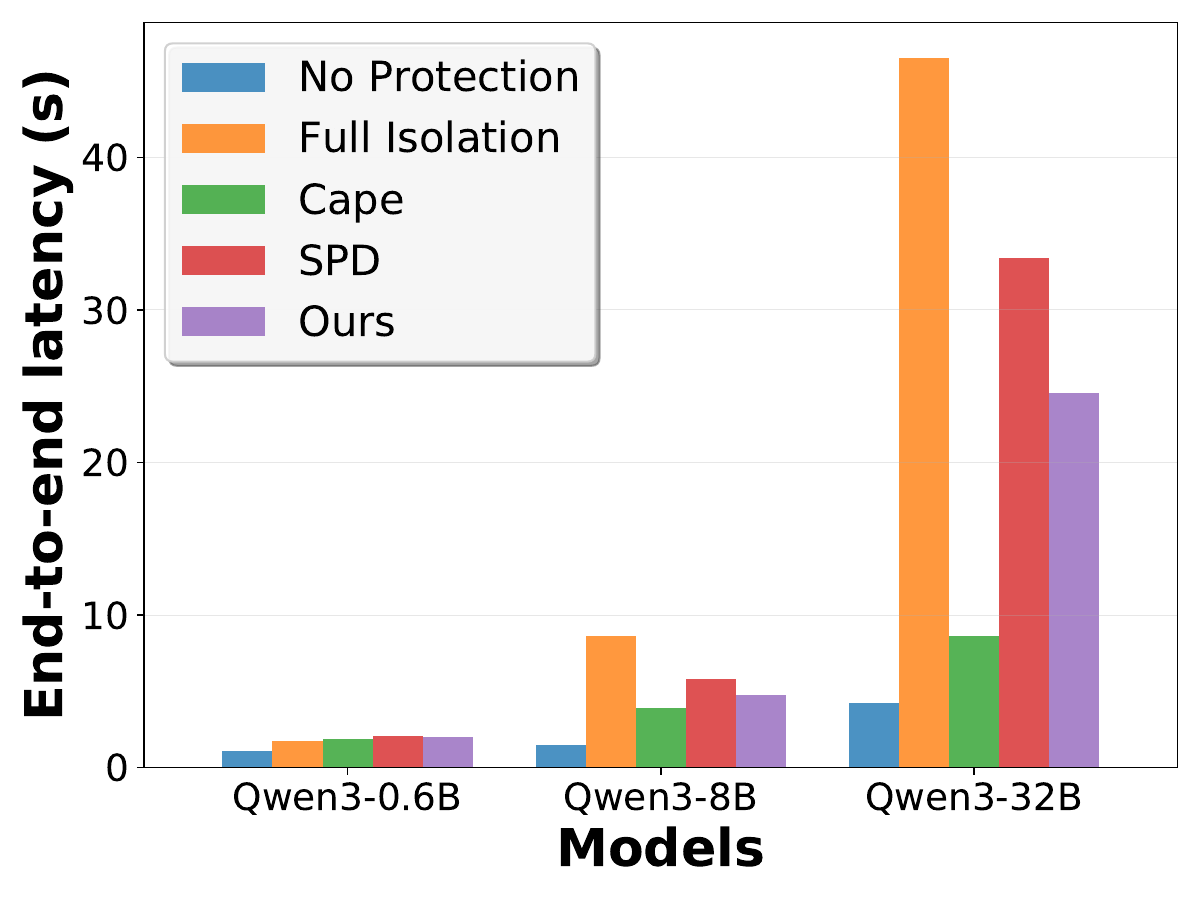}
        \caption{End-to-end latency with different model parameters. (10 clients)}
        \label{fig:param_size}
    \end{subfigure}
    \hfill
    \begin{subfigure}{0.32\linewidth}
        \centering
        \includegraphics[width=\linewidth]{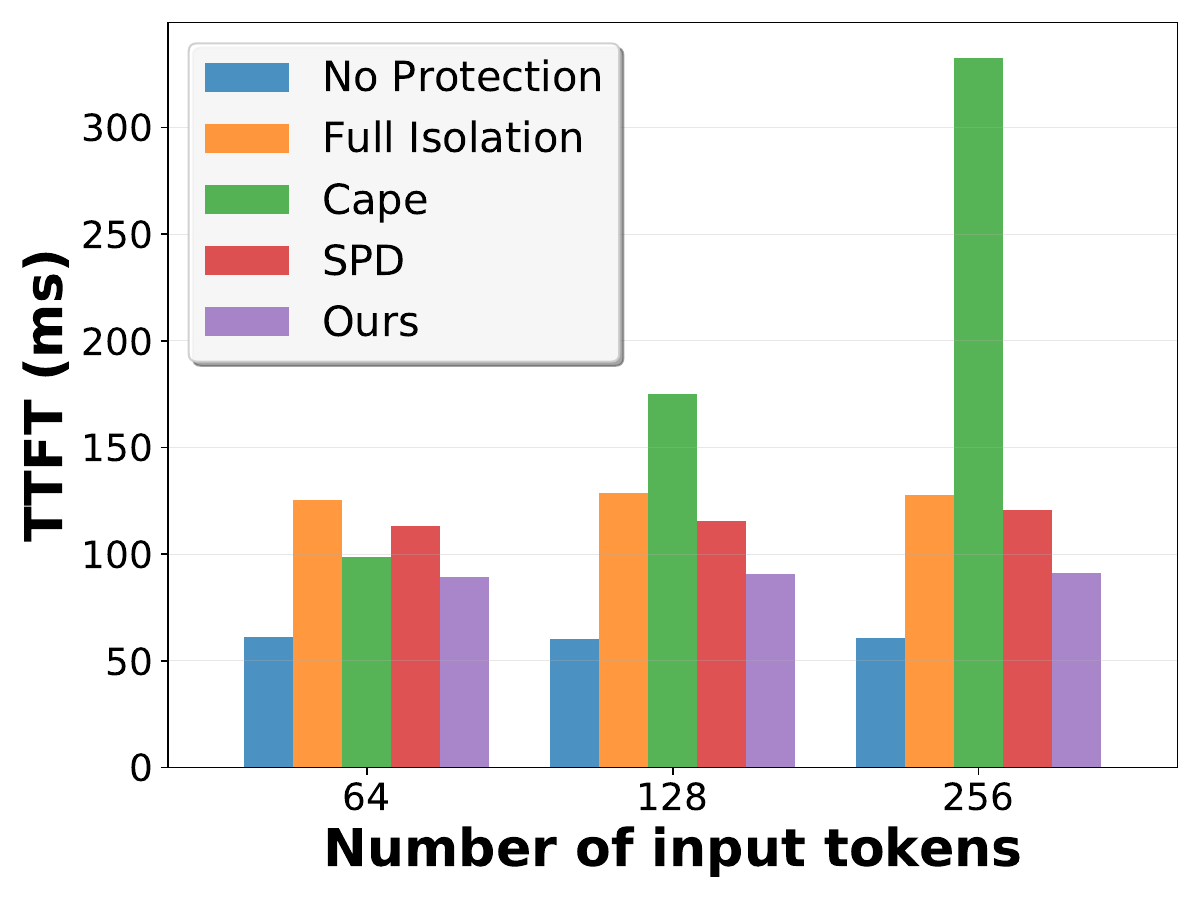}
        \caption{TTFT with varying number of input tokens. (10 clients)}
        \label{fig:input_token}
    \end{subfigure}
    \caption{Efficiency Evaluation.}
    \label{fig:efficiency}
\end{figure*}

\myparagraph{Number of clients.}
\textit{Full Isolation} requires a separate LLM instance for each CVM.
When the number of concurrent clients increases, the operating system and GPU driver must perform expensive context switches.
Therefore, as shown in Fig.~\ref{fig:num_of_clients}, the end-to-end latency curve of \textit{Full Isolation} rises sharply with the number of clients.
Compared with the SPD mechanism, \method achieves better scalability because the SPD mechanism needs to infer multiple virtual prompts simultaneously to protect client privacy, which increases the memory usage.
Compared to these two CVM-based approaches, \method demonstrates superior scalability by maintaining lower end-to-end latency as the number of concurrent clients grows.

\myparagraph{Size of model parameters.}
As model size grows, all methods exhibit increased end-to-end latency (Fig.~\ref{fig:param_size}).
Compared with \textit{Full Isolation} and SPD, \method is less affected by the expansion of parameter scale.
Intuitively, \method confines weight-free operations to the CVM and outsources weight-dependent multiplications efficiently, avoiding the per-client model duplication and heavy context switching costs that penalize \textit{Full Isolation}, as well as the additional overheads present in SPD.
\method maintains robust token generation speed under larger models, delivering lower latency growth than \textit{Full Isolation} and SPD as parameter scale expands.


\myparagraph{Number of input tokens.}
Fig.~\ref{fig:input_token} shows the change of TTFT as the number of input tokens increases (64$\rightarrow$256).
TTFT increases for all methods, but the growth is particularly steep for Cape, whose core routine computes contextual logits for every input token.
Unlike \textit{No Protection}, the three CVM-based designs (\textit{Full Isolation}, SPD, and \method) cannot amortize prefill via multi-client batching because each request must run inside an isolated CVM, which inherently raises TTFT as inputs lengthen.
Even under this constraint, \method consistently delivers the lowest TTFT among secure baselines and exhibits the mildest growth with respect to input length, yielding a faster first token--especially for long prompts.
\method preserves strong isolation while keeping prefill latency close to the practical minimum for CVM-based schemes, thus offering the fastest initial response among secure methods as input tokens scale.



\subsection{Output Invariance Evaluations.}
\label{sec:eval-invariance}
We assess whether \method preserves the exact functional behavior of the LLM by generating paired responses with and without \method under an identical decoding setup (same model, tokenizer, fixed random seed, and deterministic sampler).
We then compare the paired outputs using TRA and CSS, which respectively measure token-level identity and semantic equivalence.
As detailed in Tbl.~\ref{tab:invariance_eval}, \method achieves TRA=100\% and CSS=1.00 for each dataset, indicating bit-for-bit identical responses relative to \textit{No Protection}.
These results empirically validate the losslessness of ReMO and local restoration pipeline and confirm that \method preserves output quality without perturbation and then satisfies the performance of the ``impossible trinity'' by guaranteeing output invariance while adding privacy protection.
Additional invariance evaluations can be found in Appx.~\ref{appx:invariance}.

\begin{table}[H]
  \centering
  \adjustbox{max width=1.0\linewidth}{
  \begin{tabular}{lcccc}
    \toprule
    \textbf{Metrics}        & \textbf{Midjourney} & \textbf{WikiText2} & \textbf{Patient} & \textbf{GPT-samples} \\
    \midrule
    TRA & 100\% & 100\% & 100\% & 100\% \\
    CSS & 1.00  & 1.00  & 1.00  & 1.00 \\
    \bottomrule
  \end{tabular}
  }
  \caption{\label{tab:invariance_eval}The changes in response before and after using \method.}
\end{table}

\section{Conclusion}

We introduced \method, a confidential inference framework for cloud-based LLMs by partitioning the inference pipeline and leveraging CVMs and the ReMO protocol.
Empirically, \method cuts reconstruction success to near-negligible levels across models, datasets, and attacks, while preserving bit-for-bit outputs and competitive latency.
To the best of our knowledge, the first to keep both prompts and responses confidentiality while balancing model privacy, performance, and efficiency.

\section*{Limitations}
\label{sec:limitation}


The limitations of this work mainly include the following two aspects:

\myparagraph{Threat Scenario.}
%
We adopt the standard \emph{honest-but-curious} setting: the cloud provider executes the protocol correctly but may analyze observed artifacts to infer private information.
This assumption is reasonable for cloud LLM services because \ding{172} providers are economically and reputationally incentivized to follow deployed protocols, making arbitrary deviation unlikely in practice; \ding{173} confidential-computing primitives (e.g., CVM attestation and capability scoping) already constrain what either party can do while still enabling routine operations; and \ding{174} it aligns with the dominant threat model used in prior confidential-inference work~\cite{abs-2409-19134,abs-2505-05922} and in our system design.
Under this setting, our guarantees are testable, and the resulting system remains practical to deploy at scale.

We do not defend against a fully malicious or compromised cloud provider that can arbitrarily deviate from the protocol (e.g., returning malformed messages or bypassing attested images).
Defending against such adversaries would require stronger mechanisms (e.g., verifiable execution, runtime auditing, or side-channel–resilient defenses), which we leave to future work.

\myparagraph{Partial model protection.}
\method does not attempt to hide the entire model implementation from the client.
Instead, it prioritizes \emph{weight confidentiality}: proprietary weight parameters remain on the cloud side, while weight-free operators (e.g., normalization, attention scoring/softmax, residual connections, and sampling) are executed within the CVM.
This choice is deliberate and reasonable.
In contemporary LLM stacks, the structure of these operators is largely standardized and often public, whereas the learned weight matrices encode the provider's core intellectual property and present the primary model-extraction risk.
By keeping all weight-dependent computations on the cloud provider's hardware and exposing only weight-free control flow, the client may observe the computation pattern but not the parameters.

Moreover, visibility into non-weight operators does not enable parameter recovery under our protocol.
The client observes at most a fixed, underdetermined sketch \(R_{\text{pub}} = M_{\text{pub}} W\) with \(m < d\), leaving a non-trivial right kernel and thus infinitely many \(W\) consistent with the observation; without additional side information, the true weights are non-identifiable.
In contrast, fully obscuring every non-weight detail would require heavy obfuscation or cryptography, undermining our efficiency and output-invariance goals.
Prioritizing weight confidentiality therefore strikes a practical balance among privacy, performance, and scalability, yielding bit-for-bit identical outputs while protecting the provider's proprietary parameters.

\bibliography{custom}

@inproceedings{LinYMZH0WLCD025,
  author       = {Yu Lin and
                  Ruining Yang and
                  Yunlong Mao and
                  Qizhi Zhang and
                  Jue Hong and
                  Quanwei Cai and
                  Ye Wu and
                  Huiqi Liu and
                  Zhiyu Chen and
                  Bing Duan and
                  Sheng Zhong},
  title        = {ObfusLM: Privacy-preserving Language Model Service against Embedding
                  Inversion Attacks},
  booktitle    = {Proceedings of the 63rd Annual Meeting of the Association for Computational
                  Linguistics (Volume 1: Long Papers), {ACL} 2025, Vienna, Austria,
                  July 27 - August 1, 2025},
  pages        = {1160--1174},
  publisher    = {Association for Computational Linguistics},
  year         = {2025}
}

@misc{Quebec,
  author       = {{LégisQuébec}},
  title        = {Act respecting the protection of personal information in the private sector},
  year         = {2021},
  url          = {https://www.legisquebec.gouv.qc.ca/en/document/cs/p-39.1#:~:text=3,or%20destruction%20of%20personal%20information},
  note         = {Accessed: 2025-09-21}
}

@misc{ftc,
  author       = {Christopher Bissex},
  title        = {FTC Launches Inquiry into AI Chatbots Acting as Companions},
  year         = {2025},
  url          = {https://www.ftc.gov/news-events/news/press-releases/2025/09/ftc-launches-inquiry-ai-chatbots-acting-companions#:~:text=The%20Federal%20Trade%20Commission%20is,technology%20on%20children%20and%20teens},
  note         = {Accessed: 2025-09-21}
}

@misc{gdpr,
  author       = {Ben Wolford},
  title        = {Data Protection Impact Assessment (DPIA)},
  year         = {2025},
  url          = {https://gdpr.eu/data-protection-impact-assessment-template/},
  note         = {Accessed: 2025-09-21}
}

@misc{openai-fine,
  author       = {Elvira Pollina and Alvise Armellini},
  title        = {Italy fines OpenAI over ChatGPT privacy rules breach},
  year         = {2024},
  url          = {https://www.reuters.com/technology/italy-fines-openai-15-million-euros-over-privacy-rules-breach-2024-12-20/#:~:text=MILAN%2C%20Dec%2020%20%28Reuters%29%20,the%20generative%20artificial%20intelligence%20application},
  note         = {Accessed: 2025-09-21}
}

@misc{googlecloud,
  author       = {Google},
  title        = {Google Cloud Confidential Computing},
  year         = {2025},
  url          = {https://cloud.google.com/security/products/confidential-computing},
  note         = {Accessed: 2025-09-21}
}

@misc{azure,
  author       = {Azure},
  title        = {Azure Confidential GPU options},
  year         = {2025},
  url          = {https://learn.microsoft.com/en-us/azure/confidential-computing/gpu-options},
  note         = {Accessed: 2025-09-21}
}

@misc{amazon,
  author       = {AWS},
  title        = {AWS Confidential Computing},
  year         = {2025},
  url          = {https://aws.amazon.com/cn/confidential-computing/},
  note         = {Accessed: 2025-09-21}
}

@inproceedings{LoshchilovH19,
  author       = {Ilya Loshchilov and
                  Frank Hutter},
  title        = {Decoupled Weight Decay Regularization},
  booktitle    = {7th International Conference on Learning Representations, {ICLR} 2019,
                  New Orleans, LA, USA, May 6-9, 2019},
  publisher    = {OpenReview.net},
  year         = {2019},
}

@misc{Nvidia-llm-security,
  author       = {Laura Martinez},
  title        = {Advancing Security for Large Language Models with NVIDIA GPUs and Edgeless Systems},
  year         = {2024},
  url          = {https://developer.nvidia.com/blog/advancing-security-for-large-language-models-with-nvidia-gpus-and-edgeless-systems/},
  note         = {Accessed: 2025-09-21}
}

@misc{AzureAI,
  author       = {Mark Russinovich},
  title        = {Azure AI Confidential Inferencing: Technical Deep-Dive},
  year         = {2024},
  url          = {https://techcommunity.microsoft.com/blog/azureconfidentialcomputingblog/azure-ai-confidential-inferencing-technical-deep-dive/4253150},
  note         = {Accessed: 2025-09-21}
}

@article{XueCLHH18,
  author       = {Kaiping Xue and
                  Weikeng Chen and
                  Wei Li and
                  Jianan Hong and
                  Peilin Hong},
  title        = {Combining Data Owner-Side and Cloud-Side Access Control for Encrypted
                  Cloud Storage},
  journal      = {{IEEE} Trans. Inf. Forensics Secur.},
  volume       = {13},
  number       = {8},
  pages        = {2062--2074},
  year         = {2018},
}

@article{SharmaC21,
  author       = {Sagar Sharma and
                  Keke Chen},
  title        = {Confidential machine learning on untrusted platforms: a survey},
  journal      = {Cybersecur.},
  volume       = {4},
  number       = {1},
  pages        = {30},
  year         = {2021},
}

@article{SchwarzWGMM20,
  author       = {Michael Schwarz and
                  Samuel Weiser and
                  Daniel Gruss and
                  Cl{\'{e}}mentine Maurice and
                  Stefan Mangard},
  title        = {Malware Guard Extension: abusing Intel {SGX} to conceal cache attacks},
  journal      = {Cybersecur.},
  volume       = {3},
  number       = {1},
  pages        = {2},
  year         = {2020},
}

@inproceedings{SchwarzR24,
  author       = {Fabian Schwarz and
                  Christian Rossow},
  title        = {00SEVen - Re-enabling Virtual Machine Forensics: Introspecting Confidential
                  VMs Using Privileged in-VM Agents},
  booktitle    = {33rd {USENIX} Security Symposium, {USENIX} Security 2024, Philadelphia,
                  PA, USA, August 14-16, 2024},
  publisher    = {{USENIX} Association},
  year         = {2024},
}

@article{comanici2025gemini,
  author       = {Gheorghe Comanici and
                  Eric Bieber and
                  Mike Schaekermann and
                  Ice Pasupat and
                  Noveen Sachdeva and
                  Inderjit S. Dhillon and
                  Marcel Blistein and
                  Ori Ram and
                  Dan Zhang and
                  Evan Rosen and
                  Luke Marris and
                  Sam Petulla and
                  Colin Gaffney and
                  Asaf Aharoni and
                  Nathan Lintz and
                  Tiago Cardal Pais and
                  Henrik Jacobsson and
                  Idan Szpektor and
                  Nan{-}Jiang Jiang and
                  Krishna Haridasan and
                  Ahmed Omran and
                  Nikunj Saunshi and
                  Dara Bahri and
                  Gaurav Mishra and
                  Eric Chu and
                  Toby Boyd and
                  Brad Hekman and
                  Aaron Parisi and
                  Chaoyi Zhang and
                  Kornraphop Kawintiranon and
                  Tania Bedrax{-}Weiss and
                  Oliver Wang and
                  Ya Xu and
                  Ollie Purkiss and
                  Uri Mendlovic and
                  Ila{\"{\i}} Deutel and
                  Nam Nguyen and
                  Adam Langley and
                  Flip Korn and
                  Lucia Rossazza and
                  Alexandre Ram{\'{e}} and
                  Sagar Waghmare and
                  Helen Miller and
                  Nathan Byrd and
                  Ashrith Sheshan and
                  Raia Hadsell Sangnie Bhardwaj and
                  Pawel Janus and
                  Tero Rissa and
                  Dan Horgan and
                  Sharon Silver and
                  Ayzaan Wahid and
                  Sergey Brin and
                  Yves Raimond and
                  Klemen Kloboves and
                  Cindy Wang and
                  Nitesh Bharadwaj Gundavarapu and
                  Ilia Shumailov and
                  Bo Wang and
                  Mantas Pajarskas and
                  Joe Heyward and
                  Martin Nikoltchev and
                  Maciej Kula and
                  Hao Zhou and
                  Zachary Garrett and
                  Sushant Kafle and
                  Sercan Arik and
                  Ankita Goel and
                  Mingyao Yang and
                  Jiho Park and
                  Koji Kojima and
                  Parsa Mahmoudieh and
                  Koray Kavukcuoglu and
                  Grace Chen and
                  Doug Fritz and
                  Anton Bulyenov and
                  Sudeshna Roy and
                  Dimitris Paparas and
                  Hadar Shemtov and
                  Bo{-}Juen Chen and
                  Robin Strudel and
                  David Reitter and
                  Aurko Roy and
                  Andrey Vlasov and
                  Changwan Ryu and
                  Chas Leichner and
                  Haichuan Yang and
                  Zelda Mariet and
                  Denis Vnukov and
                  Tim Sohn and
                  Amy Stuart and
                  Wei Liang and
                  Minmin Chen and
                  Praynaa Rawlani and
                  Christy Koh and
                  JD Co{-}Reyes and
                  Guangda Lai and
                  Praseem Banzal and
                  Dimitrios Vytiniotis and
                  Jieru Mei and
                  Mu Cai},
  title        = {Gemini 2.5: Pushing the Frontier with Advanced Reasoning, Multimodality,
                  Long Context, and Next Generation Agentic Capabilities},
  journal      = {CoRR},
  volume       = {abs/2507.06261},
  year         = {2025},
}

@misc{gpt5,
  author       = {OpenAI},
  title        = {Gpt-5 system card},
  year         = {2025},
  url          = {https://openai.com/index/gpt-5-system-card/},
  note         = {Accessed: 2025-08-18}
}

@inproceedings{DevlinCLT19,
  author       = {Jacob Devlin and
                  Ming{-}Wei Chang and
                  Kenton Lee and
                  Kristina Toutanova},
  title        = {{BERT:} Pre-training of Deep Bidirectional Transformers for Language
                  Understanding},
  booktitle    = {Proceedings of the 2019 Conference of the North American Chapter of
                  the Association for Computational Linguistics: Human Language Technologies,
                  {NAACL-HLT} 2019, Minneapolis, MN, USA, June 2-7, 2019, Volume 1 (Long
                  and Short Papers)},
  pages        = {4171--4186},
  publisher    = {Association for Computational Linguistics},
  year         = {2019},
}

@misc{openai2025gptoss120bgptoss20bmodel,
  title         = {GPT-OSS-120b \& GPT-OSS-20B Model Card}, 
  author        = {OpenAI},
  year          = {2025},
  eprint        = {2508.10925},
  archivePrefix = {arXiv},
  primaryClass  = {cs.CL},
}

@inproceedings{VaswaniSPUJGKP17,
  author       = {Ashish Vaswani and
                  Noam Shazeer and
                  Niki Parmar and
                  Jakob Uszkoreit and
                  Llion Jones and
                  Aidan N. Gomez and
                  Lukasz Kaiser and
                  Illia Polosukhin},
  title        = {Attention is All you Need},
  booktitle    = {Advances in Neural Information Processing Systems 30: Annual Conference
                  on Neural Information Processing Systems 2017, December 4-9, 2017,
                  Long Beach, CA, {USA}},
  pages        = {5998--6008},
  year         = {2017},
}

@inproceedings{ZhangS19a,
  author       = {Biao Zhang and
                  Rico Sennrich},
  title        = {Root Mean Square Layer Normalization},
  booktitle    = {Advances in Neural Information Processing Systems 32: Annual Conference
                  on Neural Information Processing Systems 2019, NeurIPS 2019, December
                  8-14, 2019, Vancouver, BC, Canada},
  pages        = {12360--12371},
  year         = {2019},
}

@misc{patientnotes,
  author       = {Le An Ha and
                  Maggie and
                  Ryan Holbrook and
                  Victoria Yaneva},
  title        = {NBME - Score Clinical Patient Notes},
  year         = {2022},
  url          = {https://kaggle.com/competitions/nbme-score-clinical-patient-notes},
  note         = {Accessed: 2025-08-18}
}

@misc{nvidiacc,
  author    = {Nvidia},
  title     = {Nvidia confidential computing},
  year      = {2023},
  url       = {https://www.nvidia.com/en-us/data-center/solutions/confidential-computing/},
  note      = {Accessed: 2025-08-18}
}

@misc{amdsev,
  author    = {AMD},
  title     = {AMD Secure Encrypted Virtualization (SEV)},
  year      = {2023},
  url       = {https://www.amd.com/en/developer/sev.html},
  note      = {Accessed: 2025-08-18}
}

@misc{qwen3technicalreport,
  title         = {Qwen3 Technical Report}, 
  author        = {Qwen Team},
  year          = {2025},
  eprint        = {2505.09388},
  archivePrefix = {arXiv},
  primaryClass  = {cs.CL},
  url           = {https://arxiv.org/abs/2505.09388}, 
}

@article{GuKRVM23,
  author       = {Kang Gu and
                  Ehsanul Kabir and
                  Neha Ramsurrun and
                  Soroush Vosoughi and
                  Shagufta Mehnaz},
  title        = {Towards Sentence Level Inference Attack Against Pre-trained Language
                  Models},
  journal      = {Proc. Priv. Enhancing Technol.},
  volume       = {2023},
  number       = {3},
  pages        = {62--78},
  year         = {2023},
}

@inproceedings{0004ZWXYLZ25,
  author       = {Wenjie Qu and
                  Yuguang Zhou and
                  Yongji Wu and
                  Tingsong Xiao and
                  Binhang Yuan and
                  Yiming Li and
                  Jiaheng Zhang},
  title        = {Prompt Inversion Attack Against Collaborative Inference of Large Language
                  Models},
  booktitle    = {{IEEE} Symposium on Security and Privacy, {SP} 2025, San Francisco,
                  CA, USA, May 12-15, 2025},
  pages        = {1695--1712},
  publisher    = {{IEEE}},
  year         = {2025},
}

@misc{edoardo_federici_2022,
    author       = {Edoardo Federici},
    title        = {sentence-bert-base, sentence-transformer for Italian},
    year         = 2022,
    url          = {https://huggingface.co/efederici/sentence-bert-base},
    doi          = {10.57967/hf/0112},
    publisher    = {Hugging Face}
}

@inproceedings{MerityX0S17,
  author       = {Stephen Merity and
                  Caiming Xiong and
                  James Bradbury and
                  Richard Socher},
  title        = {Pointer Sentinel Mixture Models},
  booktitle    = {5th International Conference on Learning Representations, {ICLR} 2017,
                  Toulon, France, April 24-26, 2017, Conference Track Proceedings},
  publisher    = {OpenReview.net},
  year         = {2017},
}

@misc{gpt-oss20b-samples,
  author    = {Jack Morris},
  title     = {jxm/gpt-oss20b-samples},
  year      = {2025},
  url       = {https://huggingface.co/datasets/jxm/gpt-oss20b-samples},
  note      = {Accessed: 2025-08-18}
}

@misc{midjourney-prompts,
  author    = {Succinctly AI},
  title     = {Datasets: succinctly/midjourney-prompts},
  year      = {2024},
  url       = {https://huggingface.co/datasets/succinctly/midjourney-prompts},
  note      = {Accessed: 2025-08-04}
}

@misc{llama3,
  author    = {Meta},
  title     = {Llama 3},
  year      = {2025},
  url       = {https://www.llama.com/models/llama-3/},
  note      = {Accessed: 2025-08-04}
}

@misc{qwen3,
  author    = {Alibaba},
  title     = {Qwen 3},
  year      = {2025},
  url       = {https://qwenlm.github.io/blog/qwen3/},
  note      = {Accessed: 2025-08-04}
}

@article{abs-2503-09291,
  author       = {Xinjian Luo and
                  Ting Yu and
                  Xiaokui Xiao},
  title        = {Prompt Inference Attack on Distributed Large Language Model Inference
                  Frameworks},
  journal      = {CoRR},
  volume       = {abs/2503.09291},
  year         = {2025},
}

@inproceedings{LeeLPLLLXXZS19,
  author       = {Taegyeong Lee and
                  Zhiqi Lin and
                  Saumay Pushp and
                  Caihua Li and
                  Yunxin Liu and
                  Youngki Lee and
                  Fengyuan Xu and
                  Chenren Xu and
                  Lintao Zhang and
                  Junehwa Song},
  title        = {Occlumency: Privacy-preserving Remote Deep-learning Inference Using
                  {SGX}},
  booktitle    = {The 25th Annual International Conference on Mobile Computing and Networking,
                  MobiCom 2019, Los Cabos, Mexico, October 21-25, 2019},
  pages        = {46:1--46:17},
  publisher    = {{ACM}},
  year         = {2019},
}

@inproceedings{Hanzlik0G0A0F21,
  author       = {Lucjan Hanzlik and
                  Yang Zhang and
                  Kathrin Grosse and
                  Ahmed Salem and
                  Maximilian Augustin and
                  Michael Backes and
                  Mario Fritz},
  title        = {MLCapsule: Guarded Offline Deployment of Machine Learning as a Service},
  booktitle    = {{IEEE} Conference on Computer Vision and Pattern Recognition Workshops,
                  {CVPR} Workshops 2021, virtual, June 19-25, 2021},
  pages        = {3300--3309},
  publisher    = {Computer Vision Foundation / {IEEE}},
  year         = {2021},
}

@inproceedings{HongWZL0W24,
  author       = {Junyuan Hong and
                  Jiachen T. Wang and
                  Chenhui Zhang and
                  Zhangheng Li and
                  Bo Li and
                  Zhangyang Wang},
  title        = {{DP-OPT:} Make Large Language Model Your Privacy-Preserving Prompt
                  Engineer},
  booktitle    = {The Twelfth International Conference on Learning Representations,
                  {ICLR} 2024, Vienna, Austria, May 7-11, 2024},
  publisher    = {OpenReview.net},
  year         = {2024},
}

@inproceedings{Dwork06,
  author       = {Cynthia Dwork},
  title        = {Differential Privacy},
  booktitle    = {Automata, Languages and Programming, 33rd International Colloquium,
                  {ICALP} 2006, Venice, Italy, July 10-14, 2006, Proceedings, Part {II}},
  series       = {Lecture Notes in Computer Science},
  volume       = {4052},
  pages        = {1--12},
  publisher    = {Springer},
  year         = {2006},
}

@inproceedings{HuangLHD22,
  author       = {Zhicong Huang and
                  Wen{-}jie Lu and
                  Cheng Hong and
                  Jiansheng Ding},
  editor       = {Kevin R. B. Butler and
                  Kurt Thomas},
  title        = {Cheetah: Lean and Fast Secure Two-Party Deep Neural Network Inference},
  booktitle    = {31st {USENIX} Security Symposium, {USENIX} Security 2022, Boston,
                  MA, USA, August 10-12, 2022},
  pages        = {809--826},
  publisher    = {{USENIX} Association},
  year         = {2022},
}

@article{abs-2305-18396,
  author       = {Xuanqi Liu and
                  Zhuotao Liu},
  title        = {LLMs Can Understand Encrypted Prompt: Towards Privacy-Computing Friendly
                  Transformers},
  journal      = {CoRR},
  volume       = {abs/2305.18396},
  year         = {2023},
}

@article{abs-2406-01394,
  author       = {Ziqian Zeng and
                  Jianwei Wang and
                  Zhengdong Lu and
                  Huiping Zhuang and
                  Cen Chen},
  title        = {PrivacyRestore: Privacy-Preserving Inference in Large Language Models
                  via Privacy Removal and Restoration},
  journal      = {CoRR},
  volume       = {abs/2406.01394},
  year         = {2024},
}

@article{abs-2309-03057,
  author       = {Yu Chen and
                  Tingxin Li and
                  Huiming Liu and
                  Yang Yu},
  title        = {Hide and Seek (HaS): {A} Lightweight Framework for Prompt Privacy
                  Protection},
  journal      = {CoRR},
  volume       = {abs/2309.03057},
  year         = {2023},
}

@article{abs-2505-05922,
  author       = {Haoqi Wu and
                  Wei Dai and
                  Li Wang and
                  Qiang Yan},
  title        = {Cape: Context-Aware Prompt Perturbation Mechanism with Differential
                  Privacy},
  journal      = {CoRR},
  volume       = {abs/2505.05922},
  year         = {2025},
}

@inproceedings{WuPWM24,
  author       = {Tong Wu and
                  Ashwinee Panda and
                  Jiachen T. Wang and
                  Prateek Mittal},
  title        = {Privacy-Preserving In-Context Learning for Large Language Models},
  booktitle    = {The Twelfth International Conference on Learning Representations,
                  {ICLR} 2024, Vienna, Austria, May 7-11, 2024},
  publisher    = {OpenReview.net},
  year         = {2024},
}

@article{abs-2412-10652,
  author       = {Jinglong Luo and
                  Guanzhong Chen and
                  Yehong Zhang and
                  Shiyu Liu and
                  Hui Wang and
                  Yue Yu and
                  Xun Zhou and
                  Yuan Qi and
                  Zenglin Xu},
  title        = {Centaur: Bridging the Impossible Trinity of Privacy, Efficiency, and
                  Performance in Privacy-Preserving Transformer Inference},
  journal      = {CoRR},
  volume       = {abs/2412.10652},
  year         = {2024},
}

@article{abs-2409-19134,
  author       = {In Gim and
                  Caihua Li and
                  Lin Zhong},
  title        = {Confidential Prompting: Protecting User Prompts from Cloud {LLM} Providers},
  journal      = {CoRR},
  volume       = {abs/2409.19134},
  year         = {2024},
}

@inproceedings{LinHWJFZ25,
  author       = {Sam Lin and
                  Wenyue Hua and
                  Zhenting Wang and
                  Mingyu Jin and
                  Lizhou Fan and
                  Yongfeng Zhang},
  title        = {EmojiPrompt: Generative Prompt Obfuscation for Privacy-Preserving
                  Communication with Cloud-based LLMs},
  booktitle    = {Proceedings of the 2025 Conference of the Nations of the Americas
                  Chapter of the Association for Computational Linguistics: Human Language
                  Technologies, {NAACL} 2025 - Volume 1: Long Papers, Albuquerque, New
                  Mexico, USA, April 29 - May 4, 2025},
  pages        = {12342--12361},
  publisher    = {Association for Computational Linguistics},
  year         = {2025},
}

@misc{paul2023,
  author    = {Paul Farrell},
  title     = {JPMorgan restricts ChatGPT usage for its 250K staff over fears it could steal sensitive banking secrets},
  year      = {2023},
  url       = {https://www.dailymail.co.uk/news/article-11780501/JPMorgan-restricts-ChatGPT-usage-250-000-staff-data-privacy-fears.html},
  note      = {Accessed: 2025-07-02}
}

@misc{brian2023,
  author    = {Brian Bushard},
  title     = {Workers' ChatGPT Use Restricted At More Banks—Including Goldman, Citigroup},
  year      = {2023},
  url       = {https://www.forbes.com/sites/brianbushard/2023/02/24/workers-chatgpt-use-restricted-at-more-banks-including-goldman-citigroup/},
  note      = {Accessed: 2025-07-02}
}

@misc{giles2023,
  author    = {Giles Bruce},
  title     = {Iowa health system warns against using ChatGPT to draft patient letters},
  year      = {2023},
  url       = {https://www.beckershospitalreview.com/healthcare-information-technology/cybersecurity/iowa-health-system-warns-against-using-chatgpt-to-draft-patient-letters/},
  note      = {Accessed: 2025-07-02}
}

\appendix

\section{Notations}
\label{appx:notation}
We summarize all the notations used in this paper in Tbl.~\ref{tab:notation}.

\begin{table*}[ht]
\centering
\adjustbox{max width=1.0\linewidth}{
\begin{tabular}{ll}
\toprule
\textbf{Notation} & \textbf{Description} \\
\midrule
$X$ & Client prompt (token sequence). \\
$E \in \mathbb{R}^{n \times d}$ & Intermediate embedding matrix (sequence length $n$, hidden dimension $d$). \\
$W \in \mathbb{R}^{d \times d}$ & Proprietary model weight matrix (linear projection). \\
$O = EW$ & True weighted result of applying weights. \\
$\hat{E} = E + M$ & Masked embedding after additive mask $M$. \\
$\hat{O} = \hat{E}W$ & Masked weighted output computed by the cloud. \\
$M \in \mathbb{R}^{n \times d}$ & Additive mask matrix sampled per token. \\
$M_{\mathrm{pub}} \in \mathbb{R}^{m \times d}$ & Public base mask ($m < d$) generated in setup. \\
$R_{\mathrm{pub}} = M_{\mathrm{pub}} W$ & Pre-multiplied restoration pool returned to CVM. \\
$M_{\mathrm{pvt}} \in \mathbb{R}^{n \times m}$ & Ephemeral private mixing matrix, sampled per token. \\
$M = M_{\mathrm{pvt}} M_{\mathrm{pub}}$ & Full mask derived from public and private parts. \\
$R = M_{\mathrm{pvt}} R_{\mathrm{pub}}$ & Restoration term used to recover $O$. \\
$O = \hat{O} - R$ & Local recovery of the true weighted result. \\
$\mathsf{EOS}$ & End-of-sequence token for decoding termination. \\
$k$ & Number of decoding layers. \\
$\mathcal{A}$ & Probabilistic polynomial-time adversary. \\
$\mathcal{D}_0, \mathcal{D}_1$ & Two probability distributions in indistinguishability game. \\
$Z$ & Public auxiliary information (e.g., protocol transcript, $M_{\text{pub}}$). \\
$\text{Adv}(\mathcal{A})$ & Distinguishing advantage of adversary $\mathcal{A}$. \\
$\varepsilon(\kappa,q)$ & Negligible bound in computational indistinguishability, depends on security parameter $\kappa$ and query count $q$. \\
$\kappa$ & Security parameter (PRG seed length). \\
$q$ & Number of masked samples observed by adversary. \\
$\| \cdot \|_1, \| \cdot \|_\infty$ & $\ell_1$ norm, $\ell_\infty$ norm (used in bounds). \\
$\ker(M)$ & Kernel (null space) of matrix $M$, i.e., $\{v \in \mathbb{R}^d : Mv = 0\}$. \\
$D_{\mathrm{TV}}$ & Total variation distance between two distributions. \\
$\lambda$ & Range parameter of uniform mask distribution. \\
$e = \mathrm{vec}(E) \in \mathbb{R}^N$ & Vectorized embedding, $N = nd$. \\
TRA & Token Reconstruction Accuracy (attack metric). \\
CSS & Cosine Semantic Similarity (attack metric). \\
TTFT & Time to First Token (efficiency metric). \\
\bottomrule
\end{tabular}
}
\caption{\label{tab:notation}Summary of notations used in the paper.}
\end{table*}

\section{Additional Method Details}

\subsection{Pseudocode Implementation}
\label{appx:code}
The confidential inference process of \method is described as Alg.~\ref{alg}.

\begin{algorithm}[ht]
\caption{Pseudocode of \method.}
\label{alg}
\small
\begin{algorithmic}
\State \textbf{Input:} client prompt $X$.
\State \textbf{Output:} model response.
\State \textbf{Phase 1: One-time setup}
\State Initialize $M_{\mathrm{pub}}$ \hfill [CVM]
\State $R_{\mathrm{pub}} \gets M_{\mathrm{pub}} W$ \hfill [Cloud]
\State \textbf{Phase 2: Decoding}
\State $X_{\mathrm{new}} \gets X$
\While{$X_{\mathrm{new}} \ne \mathsf{EOS}$}
    \State Initialize $M_{\mathrm{pvt}}$ \hfill [CVM]
    \State $R \gets M_{\mathrm{pvt}} R_{\mathrm{pub}}$ \hfill [CVM]
    \State $M \gets M_{\mathrm{pvt}} M_{\mathrm{pub}}$ \hfill [CVM]
    \State $E_1 \gets \Call{Embedding}{X_{\mathrm{new}}}$ \hfill [CVM]
    \For{$i = 1$ \textbf{to} $k$}
        \State $\hat E_i \gets \Call{Mask}{M, E_i}$ \hfill [CVM]
        \State $\hat O_i \gets \Call{WeightedDec}{\hat E_i, W_i}$ \hfill [Cloud]
        \State $O_i \gets \Call{Recover}{\hat O_i, R}$ \hfill [CVM]
        \State $E_{i+1} \gets \Call{StructuralDec}{O_i}$ \hfill [CVM]
    \EndFor
    \State $X_{\mathrm{new}} \gets \Call{Sample}{E_k}$ \hfill [CVM]
\EndWhile
\end{algorithmic}
\end{algorithm}

\subsection{Theoretical Analysis of Def.~\ref{def:ci}}
\label{appx:proof-def}

This section provides a theoretical foundation for Def.~\ref{def:ci}, which establishes the privacy guarantee of \method based on the standard cryptographic concept of computational indistinguishability.
This principle is a cornerstone of modern cryptography, used to formally define the security of primitives such as pseudorandom generators and encryption schemes.

\myparagraph{The indistinguishability game.}
The notion of computational indistinguishability is captured via a security game between a challenger and a Probabilistic Polynomial-Time (PPT) adversary $\mathcal{A}$:
\begin{enumerate}
    \item \textbf{Setup:} The game is defined over two probability distributions, $\mathcal{D}_0$ and $\mathcal{D}_1$. The adversary, $\mathcal{A}$, is aware of these two distributions.
    \item \textbf{Challenge:} The challenger secretly flips a fair coin, choosing a bit $b \in \{0, 1\}$. It then samples a value $v$ from the distribution $\mathcal{D}_b$ (i.e., $v \sim \mathcal{D}_b$) and sends $v$ to the adversary.
    \item \textbf{Adversary's Goal:} Upon receiving $v$, the adversary $\mathcal{A}$ must guess the original bit $b$. It outputs a guess, $b'$.
\end{enumerate}

The adversary wins the game if its guess is correct ($b' = b$). An adversary with no information can do no better than guessing randomly, succeeding with a probability of $1/2$.
The \textbf{advantage} of an adversary is defined as how much better than random guessing it can perform.
\begin{equation}
    \text{Adv}(\mathcal{A}) = \left| \Pr[b' = b] - \frac{1}{2} \right| \notag
\end{equation}
The two distributions, $\mathcal{D}_0$ and $\mathcal{D}_1$, are said to be computationally indistinguishable if the advantage of any PPT adversary is \textit{negligible}--that is, smaller than any inverse polynomial in the security parameter.

The formula presented in Def.~\ref{def:ci} is an equivalent and widely used formulation of this concept.
Let $\mathcal{A}(v)$ be the output of the adversary on input $v$.
We can normalize the adversary's output to be either 0 or 1.
Let's say $\mathcal{A}$ outputs 1 when it guesses the sample came from $\mathcal{D}_1$, and 0 otherwise.
The probability of the adversary guessing correctly can be expanded as:
\begin{align}
    \Pr[b' = b] 
    &= \Pr[b' = 1 \mid b = 1]\Pr[b=1] \notag \\
    &\hphantom{=} \quad + \Pr[b' = 0 \mid b = 0]\Pr[b=0] \notag \\
    &= \tfrac{1}{2} \Bigl( 
        \Pr[\mathcal{A}(v,Z)=1 \mid v \sim \mathcal{D}_1] \notag \\
    &\hphantom{=} \quad + \Pr[\mathcal{A}(v,Z)=0 \mid v \sim \mathcal{D}_0] 
    \Bigr) \notag \\
    &= \tfrac{1}{2} \Bigl(
        \Pr_{v \sim \mathcal{D}_1}[\mathcal{A}(v,Z)=1] \notag \\
    &\hphantom{=} \quad + \bigl(1 - \Pr_{v \sim \mathcal{D}_0}[\mathcal{A}(v,Z)=1]\bigr) 
    \Bigr) \notag
\end{align}
Now, we can substitute this into the advantage formula:
\begin{align}
    \text{Adv}(\mathcal{A}) 
    &= \Biggl| \tfrac{1}{2} \Bigl( 
        \Pr_{v \sim \mathcal{D}_1}[\mathcal{A}(v,Z)=1] \notag \\
       &\hphantom{=} \quad + 1  - \Pr_{v \sim \mathcal{D}_0}[\mathcal{A}(v,Z)=1] 
    \Bigr) - \tfrac{1}{2} \Biggr| \notag \\
    &= \Biggl| \tfrac{1}{2} 
        \Pr_{v \sim \mathcal{D}_1}[\mathcal{A}(v,Z)=1] \notag \\
        &\hphantom{=} \quad - \tfrac{1}{2} 
        \Pr_{v \sim \mathcal{D}_0}[\mathcal{A}(v,Z)=1] 
    \Biggr| \notag \\
    &= \tfrac{1}{2} \Bigl| 
        \Pr_{v \sim \mathcal{D}_1}[\mathcal{A}(v,Z)=1] \notag \\
        &\hphantom{=} \quad - \Pr_{v \sim \mathcal{D}_0}[\mathcal{A}(v,Z)=1] 
    \Bigr| \notag
\end{align}
The formula in Def.~\ref{def:ci} is:
\begin{equation}
    |\Pr[\mathcal{A}(\hat{E}_{1},Z)=1] - \Pr[\mathcal{A}(\hat{E}_{2},Z)=1]| < \epsilon \notag
\end{equation}
Here, $\hat{E}_{1}$ is a sample from the distribution of masked embeddings derived from input $E_1$, and $\hat{E}_{2}$ is a sample from the distribution derived from input $E_2$.
This expression is exactly twice the adversary's advantage.
Since a negligible function multiplied by a constant remains negligible, this formulation is cryptographically equivalent.
It directly measures the difference in the adversary's behavior when presented with samples from the two worlds, which is the essence of distinguishability.

\myparagraph{Instantiation for \method.}
In our setting, the two distributions correspond to masked embeddings:
\begin{itemize}
    \item $\mathcal{D}_1$: $\hat{E}_1 = \mathsf{Mask}(E_1)$,
    \item $\mathcal{D}_2$: $\hat{E}_2 = \mathsf{Mask}(E_2)$,
\end{itemize}
where $\mathsf{Mask}(\cdot)$ is a randomized masking transformation applied to the intermediate embedding matrix $E$.
To ensure soundness, the mask must satisfy two conditions:
\begin{enumerate}
    \item \textbf{Freshness:} A fresh independent mask is sampled for each instance (e.g., per token).
    \item \textbf{Full-rank randomness:} The distribution of masks must not confine noise to a publicly known low-dimensional subspace.
    Otherwise, the adversary could project onto the orthogonal complement and recover information about $E$ exactly.
    To prevent this leakage, the mask must span the full ambient space, e.g., via additive pads sampled from a pseudorandom generator (PRG) over a fixed modulus or via two-sided linear masking with secret invertible matrices.
\end{enumerate}
In all cases, masking is performed element-wise over a fixed ring (e.g., $\mathbb{Z}_{2^k}$ after deterministic quantization of $E$), so that arithmetic and indistinguishability are defined over a finite domain.

\myparagraph{Security guarantee.}
Formally, Def.~\ref{def:ci} states that for all PPT adversaries $\mathcal{A}$, for any two inputs $E_1,E_2$, and for any public auxiliary information $Z$ (including protocol transcripts and public parameters),
\begin{equation}
    \big|\Pr[\mathcal{A}(\hat{E}_1,Z)=1] - \Pr[\mathcal{A}(\hat{E}_2,Z)=1]\big| \leq \varepsilon(\kappa,q) \notag
\end{equation}
where $\kappa$ is the security parameter (e.g., PRG key length), $q$ is the number of masked samples observed by $\mathcal{A}$ (including adaptive/chosen-input interactions), and $\varepsilon(\kappa,q)$ is negligible in $\kappa$ for any polynomially bounded $q$.

This guarantee holds under the assumption that, conditioned on $Z$, the mask distribution is computationally indistinguishable from uniform over the chosen domain, i.e., masks are generated by a secure PRG keyed within the client’s trusted environment.

\myparagraph{Interpretation.}
Intuitively, the masking acts like a one-time pad over the chosen ring: once an embedding $E$ is combined with fresh pseudorandom noise that spans the full space, the resulting $\hat{E}$ leaks negligible information about $E$.
Therefore, even a PPT cloud adversary observing $\hat{E}$ together with $Z$ cannot distinguish whether the input embedding originated from $E_1$ or $E_2$ with non-negligible advantage.
This formalizes the confidentiality guarantee of \method.

\subsection{Theoretical Analysis of Thm.~\ref{thm:its}}
\label{appx:proof-thm}

This section provides a formal derivation for the information-theoretic security bound presented in Thm.~\ref{thm:its}.
The theorem quantifies the maximum probability with which an adversary can distinguish between two inputs, $E_1$ and $E_2$, given only their masked versions.
The analysis is ``information-theoretic'' because the resulting bound does not depend on the adversary's computational power, but only on the statistical properties of the data and the masking mechanism.

The core of the proof relies on bounding the \textbf{total variation distance ($D_{TV}$)} between the probability distributions of the two possible masked outputs.
The adversary's optimal advantage in any distinguishing task is directly related to this statistical distance.

\myparagraph{Preliminaries.}
We consider a game between a challenger and an adversary $\mathcal{A}$.
The challenger chooses one of two inputs, $E_1$ or $E_2$, with equal probability ($\frac{1}{2}$).
It then applies the random masking mechanism to produce a masked output $\hat{E} = E_b + M$, where $b \in \{1, 2\}$ and $M$ is a random mask. The adversary observes $\hat{E}$ and must guess the original bit $b$.

The optimal adversary's probability of guessing correctly is given by:
\begin{equation}
    \Pr[\mathcal{A} \text{ succeeds}] = \frac{1}{2} + \frac{1}{2} D_{TV}(P_{\hat{E}_1}, P_{\hat{E}_2}) \notag
\end{equation}
where $P_{\hat{E}_1}$ and $P_{\hat{E}_2}$ are the probability distributions of the masked outputs corresponding to inputs $E_1$ and $E_2$, respectively.
The total variation distance for two continuous distributions with probability density functions (PDFs) $p_1(x)$ and $p_2(x)$ is defined as:
\begin{equation}
    D_{TV}(P_1, P_2) = \frac{1}{2} \int_{-\infty}^{\infty} |p_1(x) - p_2(x)| \,dx \notag
\end{equation}
Our goal is to bound this distance for our additive masking scheme.

\myparagraph{Derivation for a single dimension.}
To build intuition, let us first consider a single-dimensional case.
Let the inputs be two scalars $e_1, e_2 \in \mathbb{R}$.
The mask $m$ is a random variable.
For analytical clarity, we model the mask as being drawn from a continuous uniform distribution over the range $[-\lambda/2, \lambda/2]$.
The PDF of the mask, $p_m(z)$, is:
\begin{equation}
    p_m(z) = 
    \begin{cases} 
    \frac{1}{\lambda} & \text{if } z \in [-\lambda/2, \lambda/2] \\
    0 & \text{otherwise}
    \end{cases}
    \notag
\end{equation}
The adversary observes $\hat{e} = e_b + m$. The PDF of the observed value $\hat{e}$, conditioned on the original input being $e_1$ or $e_2$, is simply a shifted version of the mask's PDF:
\begin{equation}
    p_1(x) = p_m(x - e_1) \,\,\, \text{and} \,\,\, p_2(x) = p_m(x - e_2) \notag
\end{equation}
We can now compute the total variation distance:
\begin{align}
    &D_{TV}(P_{\hat{e}_1}, P_{\hat{e}_2}) \notag \\
    = \frac{1}{2} \int_{-\infty}^{\infty} &|p_m(x - e_1) - p_m(x - e_2)| \,dx \notag
\end{align}
The functions $p_m(x - e_1)$ and $p_m(x - e_2)$ represent two rectangular windows of height $1/\lambda$ and width $\lambda$, centered at $e_1$ and $e_2$.
The integral computes half the area of the symmetric difference between these two windows.

The area of this symmetric difference is the sum of the areas of the two non-overlapping parts.
Each of these parts has a width of $|e_1 - e_2|$ and a height of $1/\lambda$, provided that $|e_1 - e_2| \le \lambda$ (i.e., the windows overlap).
The total area of the symmetric difference is $2 \times |e_1 - e_2| \times \frac{1}{\lambda}$.

Substituting this into the $D_{TV}$ formula:
\begin{align}
    D_{TV}(P_{\hat{e}_1}, P_{\hat{e}_2}) &= \min\left(\frac{1}{2} \left( 2 \cdot \frac{|e_1 - e_2|}{\lambda} \right),1\right) \notag \\ 
    &= \min \left(\frac{|e_1 - e_2|}{\lambda},1 \right) \notag
\end{align}
Thus if $|e_1-e_2| \le \lambda$, the distance scales linearly; if $|e_1-e_2|>\lambda$, the distributions are disjoint and $D_{TV}=1$.

\myparagraph{Generalization to multiple dimensions.}
Now let $E \in \mathbb{R}^{n\times d}$ and vectorize it as $\mathbf e \in \mathbb{R}^N$ with $N=nd$.
Assume that the mask $M$ has independent coordinates $m_i \sim \mathrm{Unif}[-\lambda/2,\lambda/2]$.
Then the masked outputs $\hat{\mathbf e}_1$ and $\hat{\mathbf e}_2$ are uniform over two $N$-dimensional boxes of width $\lambda$ but shifted by $\delta = \mathbf e_1-\mathbf e_2$.
The total variation distance factorizes across coordinates:
\begin{equation}
    D_{TV}(P_{\hat{\mathbf e}_1},P_{\hat{\mathbf e}_2})
    = 1 - \prod_{i=1}^N \max\!\left(1 - \tfrac{|\delta_i|}{\lambda},\,0\right) \notag
\end{equation}
Using the inequality $1-\prod_i (1-a_i) \le \sum_i a_i$ for $a_i \in [0,1]$, we obtain the upper bound:
\begin{equation}
    D_{TV}(P_{\hat{\mathbf e}_1},P_{\hat{\mathbf e}_2})
    \le \min\!\left(\frac{\|\mathbf e_1-\mathbf e_2\|_1}{\lambda},\,1\right) \notag
\end{equation}
Thus, the adversary's distinguishing advantage is controlled by the $\ell_1$ distance between the inputs relative to the masking range.

\myparagraph{Final bound.}
Substituting into the adversary’s success probability yields:
\begin{align}
    \Pr[\mathcal{A} \text{ correctly distinguishes } E_1 \text{ from } E_2 \mid \hat{E}] \notag \\
    \;\le\; \tfrac{1}{2} + \tfrac{1}{2}\min\!\left(\frac{\|E_1-E_2\|_1}{\lambda},\,1\right) \notag
\end{align}
This establishes the bound stated in Thm.~\ref{thm:its}.
The result shows that the privacy guarantee is strengthened by increasing the masking range $\lambda$, and that the adversary’s advantage over random guessing vanishes as $\lambda$ grows.

\myparagraph{Remark on assumptions.}
The above derivation holds under the assumption that the mask entries are independent uniform random variables.
If the masking distribution is generated differently (e.g., via structured or low-rank randomness), the exact form of the bound must be adapted to that distribution.
In that case, the same total variation framework applies, but one must analyze the overlap of the induced distributions accordingly.

\subsection{Theoretical Analysis of Model Confidentiality}
\label{appx:proof-model}
This section provides a formal analysis of the mathematical guarantees that underpin \method's model confidentiality, as introduced in the main paper.
The core principle is to prevent a curious client, who controls a CVM instance, from reverse-engineering the cloud's proprietary model weights, $W$.
This protection is achieved by structuring the interaction between the CVM and the public GPU as an underdetermined system of linear equations, making the unique recovery of $W$ information-theoretically impossible.

\myparagraph{Problem formulation.}
During the one-time setup phase, the CVM receives a transformed version of the weight matrix:
\begin{equation}
    R_{\mathrm{pub}} = M_{\mathrm{pub}} W \notag
\end{equation}
where $M_{\mathrm{pub}} \in \mathbb{R}^{m \times d}$ is a publicly generated base mask, $R_{\mathrm{pub}} \in \mathbb{R}^{m \times d}$ is the published sketch, and $W \in \mathbb{R}^{d \times d}$ are the secret model weights. 
From the client's perspective, $M_{\mathrm{pub}}$ and $R_{\mathrm{pub}}$ are known, while $W$ is unknown.

\myparagraph{Analysis as a system of linear equations.}
The matrix equation above can be decomposed into $d$ independent systems of linear equations, one for each column of $W$.
Let $w_j$ and $r_j$ be the $j$-th columns of $W$ and $R_{\mathrm{pub}}$, respectively. For each column $j \in \{1, \dots, d\}$, we have the following system:
\begin{equation}
    M_{\mathrm{pub}} w_j = r_j \notag
\end{equation}
In this system, the matrix $M_{\mathrm{pub}}$ is the coefficient matrix, $w_j$ is the vector of $d$ unknown variables (the weights in that column), and $r_j$ is the known result vector.

\myparagraph{Underdetermined constraint.}
The protocol enforces $m < d$, so each system $M_{\mathrm{pub}} w_j = r_j$ is underdetermined. 
Since $\mathrm{rank}(M_{\mathrm{pub}}) \le m < d$, the null space is non-trivial, with
\begin{equation}
    \dim(\ker(M_{\mathrm{pub}})) \ge d - m > 0 \notag
\end{equation}
The notation $\ker(M_{\mathrm{pub}})$ denotes the kernel (or null space) of a matrix $M_{\mathrm{pub}}$, defined as
\begin{equation}
    \ker(M_{\mathrm{pub}}) = \{ x \in \mathbb{R}^d : M_{\mathrm{pub}} x = 0 \} \notag
\end{equation}
It represents the set of directions that are invisible to $M_{\mathrm{pub}}$. 
If $\ker(M_{\mathrm{pub}})$ is non-trivial, then any solution to $M_{\mathrm{pub}}x=b$ can be shifted by an element of $\ker(M_{\mathrm{pub}})$ to produce another valid solution, yielding infinitely many possibilities.

\myparagraph{Solution space.}
Each solution set has the affine form:
\begin{equation}
    w_j = w_p + w_h \notag
\end{equation}
where $w_p$ is a particular solution (e.g., $w_p = M_{\mathrm{pub}}^+ r_j$ with the Moore–Penrose pseudoinverse), and $w_h \in \ker(M_{\mathrm{pub}})$. 
Thus, infinitely many candidate weight vectors are consistent with the observed data.

\paragraph{Proposition (Non-identifiability under no priors).}
Let $\mathcal{S}_j = \{ w_p + z : z \in \ker(M_{\mathrm{pub}}) \}$ be the solution set for column $w_j$. 
If no auxiliary constraints on $w_j$ are available (i.e., the admissible set is $\mathbb{R}^d$), then $|\mathcal{S}_j| = \infty$ and the true weight vector is not uniquely identifiable. 
Uniqueness may arise only if additional side information restricts $\mathcal{S}_j$ to a singleton (e.g., sparsity, low-rank priors, quantization).

\paragraph{Protocol constraints.}
To preserve confidentiality in practice, the following requirements are critical:
\begin{enumerate}
    \item \textbf{Fixed sketch per matrix.} For each $W$, the cloud publishes a single $R_{\mathrm{pub}} = M_{\mathrm{pub}}W$ with fixed $M_{\mathrm{pub}}$. Re-issuing with linearly independent $M_{\mathrm{pub}}$ is disallowed, since stacking multiple sketches could collapse the null space.
    \item \textbf{CVM attestation and isolation.} Although the CVM performs unmasking internally, the client cannot read intermediate tensors $(E,O=EW)$; otherwise, by collecting enough input–output pairs, the client could solve $EW=O$ and reconstruct $W$. \method relies on cloud-based attestation and runtime isolation to prevent such leakage.
    \item \textbf{Independent masking.} Distinct, untied weight matrices (e.g., $W_Q,W_K,W_V,W_O$) must use independent masks $M_{\mathrm{pub}}$ to avoid cross-module linkability.
    \item \textbf{Domain specification.} The analysis assumes real-valued arithmetic. If weights are represented in finite rings (e.g., $\mathbb{Z}_{2^k}$), rank and null-space properties must be verified in that algebraic domain.
\end{enumerate}

\paragraph{Conclusion.}
The dimensionality constraint $m < d$ guarantees that $M_{\mathrm{pub}}W$ alone does not uniquely determine $W$, ensuring non-identifiability under standard assumptions. 
Model confidentiality thus holds provided that (i) no additional independent sketches are revealed, (ii) clients cannot access intermediate tensors, and (iii) no strong structural priors on $W$ are externally available.

\section{Additional Evaluations}

\subsection{Detailed Evaluation Setup}
\label{appx:setup}

\subsubsection{Datasets}
\label{appx:dataset}
We use four representative datasets in our experiments:
\begin{itemize}
    \item Midjourney prompts~\cite{midjourney-prompts} is a dataset containing 246,381 natural language prompts for text-to-image service.
    \item WikiText-2~\cite{MerityX0S17} contains over 100 million tokens extracted from high-quality, curated Wikipedia articles.
    \item Patient-notes~\cite{patientnotes} is a dataset containing 40,000 patient note history portions.
    \item GPT-OSS20B-samples~\cite{gpt-oss20b-samples} dataset is a collection of 8,960,000 text samples generated from a version of OpenAI's GPT-OSS20B model~\cite{openai2025gptoss120bgptoss20bmodel}.
\end{itemize}

\subsubsection{Metrics}
\label{appx:metrics}

For defense evaluation, we use Token Reconstruction Accuracy (TRA) and Cosine Semantic Similarity (CSS).
TRA is formally defined as $\mathit{TRA}=\frac{1}{|\mathcal T^{tgt}|}\sum_{t_i\in \mathcal T^{tgt}}\mathbb I(\hat t_i=t_i)$ where $\mathbb I$ is the indicator function, $\mathcal T^{tgt}$ represents the set of target tokens, and $\hat t_i$ denotes the reconstructed token corresponding to $t_i$.
CSS first extracts semantic embeddings of both reconstructed and ground truth prompts using Sentence-BERT~\cite{edoardo_federici_2022} and then calculates their cosine similarity.
Based on the prior work~\cite{abs-2503-09291}, a TRA greater than 40\% is considered sufficient for plausible reconstruction.
For efficiency evaluation, we use end-to-end latency and Time to First Token (TTFT).
End-to-end latency measures the end-to-end time from when the client sends a prompt to when they receive the complete response.
Time to first token measures the time from request arrival to the generation of the first output token.

\begin{table*}[ht]
  \centering
  \adjustbox{max width=1.0\linewidth}{
  \begin{tabular}{clccccccccc}
    \toprule
    \textbf{Attack Target} & \multicolumn{2}{c}{\textbf{Models}} & \multicolumn{2}{c}{\textbf{Midjourney}} & \multicolumn{2}{c}{\textbf{WikiText2}} & \multicolumn{2}{c}{\textbf{Patient}} & \multicolumn{2}{c}{\textbf{GPT-samples}} \\
    \midrule
    \multirow{6}{*}{Prompt} 
      & \multirow{3}{*}{Qwen3}  & 0.6B & \textbf{1.56\%} & \textcolor{midgray}{0.4573} & \textbf{0.89\%} & \textcolor{midgray}{0.4453} & \textbf{1.06\%} & \textcolor{midgray}{0.4835} & \textbf{1.88\%} & \textcolor{midgray}{0.4735} \\
      & & 8B   & \textbf{1.24\%} & \textcolor{midgray}{0.4876} & 
                 \textbf{0.47\%} & \textcolor{midgray}{0.4299} & 
                 \textbf{1.53\%} & \textcolor{midgray}{0.5012} & 
                 \textbf{2.11\%} & \textcolor{midgray}{0.4535} \\
      & & 32B  & \textbf{1.43\%} & \textcolor{midgray}{0.4753} & \textbf{1.06\%} & \textcolor{midgray}{0.5022} & \textbf{1.25\%} & \textcolor{midgray}{0.4875} & \textbf{1.63\%} & \textcolor{midgray}{0.4925} \\ 
      \cmidrule(l){2-11}
      & \multirow{3}{*}{Llama3} & 1B   & \textbf{1.46\%} & \textcolor{midgray}{0.4323} & \textbf{0.93\%} & \textcolor{midgray}{0.4687} & \textbf{1.14\%} & \textcolor{midgray}{0.4566} & \textbf{1.88\%} & \textcolor{midgray}{0.4753}  \\
      & & 8B   & \textbf{1.39\%} & \textcolor{midgray}{0.4632} & \textbf{0.56\%} & \textcolor{midgray}{0.4436} & \textbf{1.42\%} & \textcolor{midgray}{0.4653} & \textbf{1.68\%} & \textcolor{midgray}{0.4923} \\
      & & 70B  & \textbf{1.57\%} & \textcolor{midgray}{0.4553} & \textbf{0.88\%} & \textcolor{midgray}{0.4329} & \textbf{1.32\%} & \textcolor{midgray}{0.4587} & \textbf{1.79\%} & \textcolor{midgray}{0.4780} \\
    \midrule
    \multirow{6}{*}{Response} 
      & \multirow{3}{*}{Qwen3}  & 0.6B & \textbf{1.34\%} & \textcolor{midgray}{0.4326} & \textbf{0.42\%} & \textcolor{midgray}{0.4420} & \textbf{0.96\%} & \textcolor{midgray}{0.4885} & \textbf{1.97\%} & \textcolor{midgray}{0.4821} \\
      & & 8B   & \textbf{1.08\%} & \textcolor{midgray}{0.4724} & 
                 \textbf{0.38\%} & \textcolor{midgray}{0.4242} & 
                 \textbf{1.58\%} & \textcolor{midgray}{0.4980} & 
                 \textbf{1.97\%} & \textcolor{midgray}{0.4463} \\
      & & 32B  & \textbf{1.21\%} & \textcolor{midgray}{0.4683} & \textbf{0.43\%} & \textcolor{midgray}{0.4876} & \textbf{0.89\%} & \textcolor{midgray}{0.5342} & \textbf{1.48\%} & \textcolor{midgray}{0.4983} \\ 
      \cmidrule(l){2-11}
      & \multirow{3}{*}{Llama3} & 1B   & \textbf{0.89\%} & \textcolor{midgray}{0.4521} & \textbf{0.87\%} & \textcolor{midgray}{0.4853} & \textbf{0.98\%} & \textcolor{midgray}{0.5421} & \textbf{1.28\%} & \textcolor{midgray}{0.5021}  \\
      & & 8B   & \textbf{1.15\%} & \textcolor{midgray}{0.4646} & \textbf{0.38\%} & \textcolor{midgray}{0.4154} & \textbf{0.76\%} & \textcolor{midgray}{0.5015} & \textbf{1.54\%} & \textcolor{midgray}{0.4875} \\
      & & 70B  & \textbf{0.78\%} & \textcolor{midgray}{0.4328} & \textbf{0.25\%} & \textcolor{midgray}{0.4423} & \textbf{0.92\%} & \textcolor{midgray}{0.5201} & \textbf{1.29\%} & \textcolor{midgray}{0.4857} \\
    \bottomrule
  \end{tabular}
  }
  \caption{\label{tab:protection_model_eval_full}Effectiveness of \method against TokenInfer attack under different model structures and sizes. 
  Bold: TRA. Gray: CSS}
\end{table*}

\subsubsection{Attacks}
\label{appx:attack}

We validate our defense against three SOTA reconstruction attacks.

\myparagraph{Classification-based attack.}
TokenInfer~\cite{abs-2503-09291} reconstructs private tokens by intercepting the intermediate embeddings exchanged between devices in a distributed LLM framework.
The core of the attack relies on the discovery that these intermediate embeddings form distinct clusters corresponding to their input tokens.
The adversary trains a classification model to map the intercepted embeddings back to their original tokens.
The attack model is a multi-layer perceptron consisting of six layers.
Layer normalization and ReLU activation functions are used between consecutive layers.
The training epochs for the attack model are 10.

\myparagraph{Optimization-based attack.}
TokenInv~\cite{0004ZWXYLZ25} is divided into two stages:
First, a constrained optimization phase is used to recover a continuous embedding that is a close approximation of the original prompt.
Second, an adaptive discretization phase accurately converts this embedding back into discrete text tokens by using the activation data and semantic speculation from another LLM to improve accuracy and readability.
The attack model is set as optimization learning rate to 0.1, constraint coefficient to 0.1, optimization iteration number to 1500, embedding-based candidate size to 10, and semantic-based candidate size to 10.

\myparagraph{Generation-based attack.}
SentenceInv~\cite{GuKRVM23} trains a decoder model using a publicly available dataset from the same domain as a private target dataset.
The decoder learns to reconstruct the original text from its sentence embeddings, which are generated by a pre-trained language model like BERT~\cite{DevlinCLT19}.
The adversary then uses this trained decoder on the private embeddings to reconstruct the original, sensitive sentences, revealing high-level semantic information.
The attack model is an adversarial decoder, which inherits from the transformer architecture with 6 decoding layers and 8 heads.
The decoder is trained for 100 epochs with cross-entropy loss, AdamW optimizer~\cite{LoshchilovH19}, and a learning rate of 1e-4.

\begin{table*}[ht]
  \centering
  \adjustbox{max width=1.0\linewidth}{
  \begin{tabular}{clcccccccc}
    \toprule
    \textbf{Attack Target} & \textbf{Attacks} & \multicolumn{2}{c}{\textbf{Midjourney}} & \multicolumn{2}{c}{\textbf{WikiText2}} & \multicolumn{2}{c}{\textbf{Patient}} & \multicolumn{2}{c}{\textbf{GPT-samples}} \\
    \midrule
    \multirow{3}{*}{Prompt} 
      & TokenInfer  & \textbf{1.24\%} & \textcolor{midgray}{0.4876} & \textbf{0.47\%} & \textcolor{midgray}{0.4299} & \textbf{1.53\%} & \textcolor{midgray}{0.5012} & \textbf{2.11\%} & \textcolor{midgray}{0.4535} \\
      & TokenInv    & \textbf{0.45\%} & \textcolor{midgray}{0.3893} & \textbf{0.52\%} & \textcolor{midgray}{0.4021} & \textbf{0.64\%} & \textcolor{midgray}{0.3988} & \textbf{0.58\%} & \textcolor{midgray}{0.4085} \\
      & SentenceInv & -- & \textcolor{midgray}{0.4435} & -- & \textcolor{midgray}{0.3735} & -- & \textcolor{midgray}{0.4051} & -- & \textcolor{midgray}{0.3835} \\
    \midrule
    \multirow{3}{*}{Response} 
      & TokenInfer  & \textbf{1.08\%} & \textcolor{midgray}{0.4724} & \textbf{0.38\%} & \textcolor{midgray}{0.4242} & \textbf{1.58\%} & \textcolor{midgray}{0.4980} & \textbf{1.97\%} & \textcolor{midgray}{0.4463} \\
      & TokenInv    & \textbf{0.56\%} & \textcolor{midgray}{0.3954} & \textbf{0.38\%} & \textcolor{midgray}{0.3785} & \textbf{0.56\%} & \textcolor{midgray}{0.4123} & \textbf{0.72\%} & \textcolor{midgray}{0.4025} \\
      & SentenceInv & -- & -- & -- & -- & -- & -- & -- & -- \\
    \bottomrule
  \end{tabular}
  }
  \caption{\label{tab:protection_attack_eval_full}Effectiveness of \method in defending against different attacks. 
  Bold: TRA. Gray: CSS.}
\end{table*}

\subsubsection{Baselines}
\label{appx:baseline}
We use two SOTA representative defense mechanisms as baselines:
\begin{itemize}
    \item SPD~\cite{abs-2409-19134} first confines the client's prompt and private data to a CVM, while splitting the computational workload with the untrusted cloud LLM provider.
    Then SPD uses prompt obfuscation to generate multiple plausible but fake ``virtual prompts'' that are processed alongside the client's real one, making it impossible for the provider to distinguish the authentic prompt from the fake ones.
    In the evaluations, we set the number of virtual prompts to 15.
    \item Cape~\cite{abs-2505-05922} protects client privacy by locally perturbing the original prompt on the client's device before it is sent to the cloud.
    It uses a hybrid utility function that considers both the context and token similarity to find suitable replacement words.
    A bucketized sampling method is then used to efficiently select a replacement token, which ensures differential privacy while trying to maintain the prompt's original meaning.
    We set the number of buckets to 50 and use Bert~\cite{DevlinCLT19} to capture contextual information.
\end{itemize}

\subsection{Additional Protection Evaluations}
\label{appx:additional_protection_eval}
\myparagraph{Different LLMs.}
We assess the scale-robustness of \method\ by evaluating four datasets--\textsc{Midjourney}, \textsc{WikiText-2}, \textsc{Patient Notes}, and \textsc{GPT-samples}--across language models with increasing parameter scale.
For each model, we measure the effectiveness of TokenInfer attack using two metrics, TRA and CSS.
As summarized in Tbl.~\ref{tab:protection_model_eval_full}, \method consistently suppresses reconstruction across all model sizes and datasets, and the magnitude of protection remains stable as model capacity grows, indicating that the defense is largely model-agnostic.

\myparagraph{Different attacks.}
We evaluate \method against three SOTA reconstruction attacks on four datasets.
As summarized in Tbl.~\ref{tab:protection_attack_eval_full}, \method consistently depresses both indicators across all datasets and attack families.
The relative protection remains stable even when attacks are tuned to their strongest published settings, suggesting that \method provides attack-agnostic robustness rather than overfitting to any single threat.

\myparagraph{Different layers.}
We evaluate the protection performance of \method against token reconstruction attacks at different layers, and use TokenInfer to attack the outsourced inputs of different layers of Qwen3-8B.
For consistent comparison, we report both TRA and CSS over the concatenated prompt–response sequence rather than treating them separately.
As summarized in Tbl.~\ref{tab:protection_layer}, across Q/K/V projections and MLP blocks at layers 1, 5, and 10 of Qwen3-8B, \method consistently collapses TokenInfer's success from near-perfect to negligible.
Before defense, early layers leak the most (e.g., Layer-1: TRA$\approx$98-99\%, CSS$\approx$0.93-0.95), and leakage naturally diminishes with depth (Layer-10: TRA$\approx$83-84\%, CSS$\approx$0.82-0.83).
After applying \method, all submodules converge to low, layer-insensitive residuals: defended TRA stays around 1.1-1.4\% (best: 1.08–1.14\%; worst: 1.43\%), and CSS falls to 0.45-0.48.
The protection is uniform across pathways (Q, K, V, MLP vary within little TRA at a given depth), indicating that hybrid masking plus partitioned execution neutralizes the principal layer-specific leakage channels rather than merely shifting them deeper.
\method turns high pre-defense leakage at shallow layers into uniformly low, sub-2\% TRA and $\le$0.48 CSS across layers and submodules, demonstrating robust protection.

\begin{table}[ht]
  \centering
  \adjustbox{max width=1.0\linewidth}{
  \begin{tabular}{cccc}
    \toprule
    \multicolumn{2}{c}{\textbf{Attack Layer}} & \textbf{TRA} & \textbf{CSS} \\
    \midrule
    \multirow{4}{*}{Layer 1} & Q   & 98.64\%$\rightarrow$1.42\% & 0.9443$\rightarrow$0.4573 \\
                             & K   & 99.13\%$\rightarrow$1.08\% & 0.9512$\rightarrow$0.4876 \\
                             & V   & 98.54\%$\rightarrow$1.26\% & 0.9337$\rightarrow$0.4753 \\
                             & MLP & 98.04\%$\rightarrow$1.17\% & 0.9375$\rightarrow$0.4805 \\
    \midrule
    \multirow{4}{*}{Layer 5} & Q   & 93.42\%$\rightarrow$1.13\% & 0.8653$\rightarrow$0.4687 \\
                             & K   & 92.88\%$\rightarrow$1.43\% & 0.8654$\rightarrow$0.4724 \\
                             & V   & 93.10\%$\rightarrow$1.36\% & 0.8634$\rightarrow$0.4653 \\
                             & MLP & 94.05\%$\rightarrow$1.23\% & 0.8723$\rightarrow$0.4564 \\
    \midrule
    \multirow{4}{*}{Layer 10} & Q   & 84.25\%$\rightarrow$1.14\% & 0.8326$\rightarrow$0.4666 \\
                              & K   & 84.48\%$\rightarrow$1.21\% & 0.8268$\rightarrow$0.4532 \\
                              & V   & 83.21\%$\rightarrow$1.09\% & 0.8251$\rightarrow$0.4483 \\
                              & MLP & 83.45\%$\rightarrow$1.12\% & 0.8265$\rightarrow$0.4503 \\
    \bottomrule
  \end{tabular}
  }
  \caption{\label{tab:protection_layer}The protection performance of \method under different model layers (Midjourney dataset).}
\end{table}

\subsection{Additional Invariance Evaluations}
\label{appx:invariance}
Tbl.~\ref{tab:invariance_text} provides three prompts together with the corresponding responses generated with and without \method.
To remove randomness, we fix the model, tokenizer, decoding policy, and random seed.
In all cases, the outputs are identical, with green shading used to mark the one-to-one correspondence. 
These examples complement the quantitative metrics and demonstrate that \method's reversible masking and local restoration preserve the model's behavior exactly, rather than only ensuring semantic similarity.
Taken together, the results in Tbl.~\ref{tab:invariance_text} confirm that \method maintains deterministic outputs while adding privacy protection, highlighting its effectiveness as a lossless mechanism.



\begin{table*}[ht]
  \centering
  \adjustbox{max width=1.0\linewidth}{
  \begin{tabular}{p{0.32\linewidth} p{0.32\linewidth} p{0.32\linewidth}}
    \toprule
    \textbf{Prompt} & \textbf{Response w/o \method} & \textbf{Response w/ \method} \\
    \midrule
    In the United States, common starlings are exempt from the Migratory Bird Treaty Act, which prohibits the taking or killing of migratory birds. No permit is required to remove nests and eggs or kill juveniles or adults. Research was undertaken in 1966 to identify a suitable avicide that would both kill common starlings and would readily be eaten by them... & First, the text says that common starlings are exempt from the Migratory Bird Treaty Act. I know that this act is a law that protects migratory birds, so if starlings are exempt, that means they aren't protected under this law. That probably means people can kill them without a permit. The text mentions that no permit is needed to remove nests, eggs, or kill both juveniles and adults. So, even though other migratory birds are protected, starlings aren't, which is why they can be targeted... & \cellcolor{midgreen}{First, the text says that common starlings are exempt from the Migratory Bird Treaty Act. I know that this act is a law that protects migratory birds, so if starlings are exempt, that means they aren't protected under this law. That probably means people can kill them without a permit. The text mentions that no permit is needed to remove nests, eggs, or kill both juveniles and adults. So, even though other migratory birds are protected, starlings aren't, which is why they can be targeted...} \\ \midrule
    Mozart had a pet common starling which could sing part of his Piano Concerto in G Major (KV. 453). He had bought it from a shop after hearing it sing a phrase from a work he wrote six weeks previously, which had not yet been performed in public. He became very attached to the bird and arranged an elaborate funeral for it when it died three years later... & So, Mozart had a starling that could sing part of his Piano Concerto in G Major (KV 453). He bought it after hearing it mimic a phrase from a piece he hadn't performed yet. The bird was a pet he was very attached to, and he had a funeral for it. There's a suggestion that his A Musical Joke (K. 522) might reflect the starling's style. Also, other starling owners note their mimicry skills, mixing phrases inappropriately, and even strangers thinking they heard a human speak. & \cellcolor{midgreen}{So, Mozart had a starling that could sing part of his Piano Concerto in G Major (KV 453). He bought it after hearing it mimic a phrase from a piece he hadn't performed yet. The bird was a pet he was very attached to, and he had a funeral for it. There's a suggestion that his A Musical Joke (K. 522) might reflect the starling's style. Also, other starling owners note their mimicry skills, mixing phrases inappropriately, and even strangers thinking they heard a human speak.} \\ \midrule
    Stephanie Madden, a 20-year-old female, has come to the doctor's office complaining of a headache. sHE REPORTS THAT SHE HAS AHD A STIFF NECK AND ha FOR PAST DAY - SHE reports that bright lights hurt her head - with general pain all over - she denies any aura preceding the headache and reports that nothing like this has happened before - she has vommitted 3+ times over the past day... & Okay, let me try to work through this case. So Stephanie is a 20-year-old female with a headache that's been going on for a day. She has a stiff neck, sensitivity to bright lights, generalized pain, and she's vomited three times. Tylenol hasn't helped. She denies any aura, which is interesting because that's a common feature in migraines. Her family history includes migraines on her mother's side and her father has HD, which I think is Huntington's disease. She's sexually active, uses condoms, and has some substance use: some alcohol and marijuana. She's not up to date on immunizations, especially the meningitis vaccine... & \cellcolor{midgreen}{Okay, let me try to work through this case. So Stephanie is a 20-year-old female with a headache that's been going on for a day. She has a stiff neck, sensitivity to bright lights, generalized pain, and she's vomited three times. Tylenol hasn't helped. She denies any aura, which is interesting because that's a common feature in migraines. Her family history includes migraines on her mother's side and her father has HD, which I think is Huntington's disease. She's sexually active, uses condoms, and has some substance use: some alcohol and marijuana. She's not up to date on immunizations, especially the meningitis vaccine.} \\
    \bottomrule
  \end{tabular}
  }
  \caption{\label{tab:invariance_text}Output of LLMs inference after using \method (green indicates text consistency).}
\end{table*}


\end{document}